%% file: main.tex
\documentclass[sigconf]{acmart}
\usepackage{xspace}
\usepackage{enumitem}
\usepackage[most]{tcolorbox}
\usepackage{soul} 
\usepackage{subfiles}

\usepackage{tabularx}  
\usepackage{colortbl}  
\usepackage{booktabs}  

\usepackage{graphicx}

\usepackage{array}
\usepackage{placeins}

\DeclareRobustCommand*{\IEEEauthorrefmark}[1]{%
  \raisebox{0pt}[0pt][0pt]{\textsuperscript{\footnotesize #1}}%
}

\AtBeginDocument{%
  }

\acmISBN{978-1-4503-XXXX-X/2018/06}

\definecolor{testColor}{HTML}{814AB7}
\definecolor{testBrown}{HTML}{6D3C1F}
\definecolor{white}{HTML}{FFFFFF}

\definecolor{purple5}{HTML}{7209B7}
\definecolor{purple4}{HTML}{580966}
\definecolor{purple3}{HTML}{770C8A}
\definecolor{purple2}{HTML}{9410AB}
\definecolor{purple1}{HTML}{C6259E}
\definecolor{pink}{HTML}{F83A90}
\definecolor{orange2}{HTML}{FA7543}
\definecolor{orange1}{HTML}{FC9D22}
\definecolor{yellow}{HTML}{FDC500}

\definecolor{lightYellow1}{HTML}{DFB64C}
\definecolor{lightYellow2}{HTML}{D6AA41}
\definecolor{pumpkin}{HTML}{EA7D3C}
\definecolor{poppyRed}{HTML}{D64844}
\definecolor{raspberryRose}{HTML}{B23972}
\definecolor{TekheletPurple}{HTML}{572F96}
\definecolor{persianIndigo}{HTML}{2F1A7F}
\definecolor{resolutionBlue}{HTML}{1B1F8D}
\definecolor{LapisBlue}{HTML}{23639F}

\definecolor{UMGreen}{HTML}{286442}
\definecolor{UMOrange}{HTML}{e77728}

\newcommand{\Code}[1]{\begin{small}\texttt{#1}\end{small}}

\newcommand{\eg}[0]{e.g., }
\newcommand{\ie}[0]{i.e., }

\newcommand{\cf}[0]{cf. }

\newcommand{\rqone}[0]{RQ1 }
\newcommand{\rqtwo}[0]{RQ2 }

\newcommand{\mysec}[1]{\smallskip \noindent \textbf{\textit{#1.}}}

\newcommand{\participantQuote}[2]{{\color{TekheletPurple}\textit{``#1''}~({#2}).}}

\newtcolorbox{keyInsightsBox}{
  enhanced,
  sharp corners,
  colback=TekheletPurple!20!white,             
  colframe=TekheletPurple!20!white,            
  boxrule=0.5pt,
  left=3pt,
  right=3pt,
  top=3pt,
  bottom=3pt,
  before skip=3pt,
  after skip=3pt,
  fonttitle=\bfseries\small,
  coltitle=black,
  attach title to upper={},
  separator sign={\ },
}

\newcommand{\hai}[0]{frequent-AI users\xspace}
\newcommand{\lai}[0]{infrequent-AI users\xspace}
\newcommand{\haic}[0]{Frequent-AI users\xspace}
\newcommand{\laic}[0]{Infrequent-AI users\xspace}
\newcommand{\hais}[0]{frequent-AI user\xspace}

\newcommand{\gait}[0]{GenAI tools\xspace}
\newcommand{\gaits}[0]{GenAI tool\xspace}

\newcommand{\gaidt}[0]{GenAI development tools\xspace}
\newcommand{\gaidts}[0]{GenAI development tool\xspace}

\newcommand{\ppp}[0]{\textit{Productivity Pressure Paradox}\xspace}

\newif\ifdraft
\drafttrue 

\setcopyright{none} 
\renewcommand\footnotetextcopyrightpermission[1]{}

\begin{document}

\title[``Maybe We Need Some More Examples:''\\Individual and Team Drivers of Developer GenAI Tool Use]{``Maybe We Need Some More Examples:''\\Individual and Team Drivers of Developer GenAI Tool Use}

\affiliation{Courtney Miller,\IEEEauthorrefmark{1} \hspace{0.2em} Rudrajit Choudhuri,\IEEEauthorrefmark{2} \hspace{0.2em} Mara Ulloa,\IEEEauthorrefmark{3} \hspace{0.2em} Sankeerti Haniyur,\IEEEauthorrefmark{4} \hspace{0.3em} Robert DeLine,\IEEEauthorrefmark{4} \hspace{0.2em} Margaret-Anne Storey,\IEEEauthorrefmark{5} \hspace{0.2em} Emerson Murphy-Hill,\IEEEauthorrefmark{4} \hspace{0.2em} Christian Bird,\IEEEauthorrefmark{4} \hspace{0.2em} Jenna L. Butler\IEEEauthorrefmark{4} 
\country{}
}

\affiliation{
\institution{ \normalsize \IEEEauthorrefmark{1}Carnegie Mellon University, PA, USA. Email: courtneymiller@cmu.edu}
\country{}
}
\affiliation{
\institution{ \normalsize \IEEEauthorrefmark{2}Oregon State University, OR, USA. Email: choudhru@oregonstate.edu}
\country{}
}
\affiliation{
\institution{ \normalsize \IEEEauthorrefmark{3}Northwestern University, IL, USA. Email: mara.ulloa@u.northwestern.edu}
\country{}
}
\affiliation{
\institution{ \normalsize \IEEEauthorrefmark{4}Microsoft, WA, USA. Email: {sahaniyur,rdeline,emerson.rex,cbird,jennbu}@microsoft.com}
\country{}
}
\affiliation{
\institution{ \normalsize \IEEEauthorrefmark{5}University of Victoria, BC, Canada. Email: mstorey@uvic.ca}
\country{}
}

\renewcommand{\shortauthors}{Miller et al.}

\begin{abstract}
Despite the widespread availability of generative AI tools in software engineering, 
developer adoption remains uneven. 
This unevenness is problematic because it hampers productivity efforts, 
frustrates management's expectations, 
and creates uncertainty around the future roles of developers. 
Through paired interviews with 54 developers across 27 teams 
-- one frequent and one infrequent user per team -- 
we demonstrate that differences in usage result primarily from 
how developers perceive the tool (as a collaborator vs. feature), 
their engagement approach (experimental vs. conservative), 
and how they respond when encountering challenges (with adaptive persistence vs. quick abandonment). 
Our findings imply that widespread organizational expectations for rapid productivity gains 
without sufficient investment in learning support creates a 
"\textit{Productivity Pressure Paradox}," undermining the very productivity benefits that motivate  adoption.
\end{abstract}

\maketitle



\vspace{-0.2cm}
\section{Introduction}
\label{sec:Introduction}

\begin{figure}[t]
    \centering
    \includegraphics[width=\linewidth]{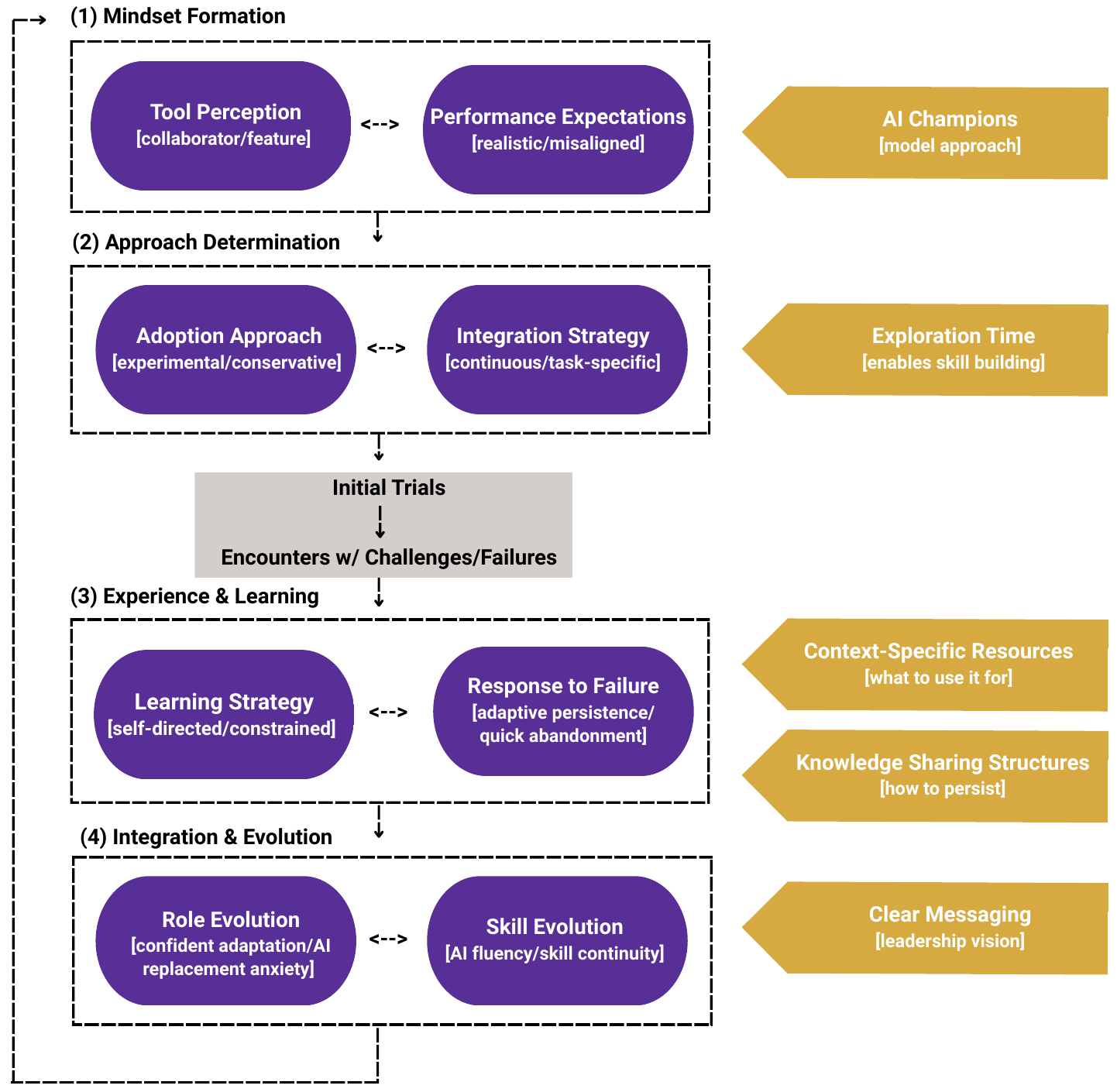}
   \vspace{-17pt}
 \caption{Conceptual theoretical framework. Purple ovals == individual factors, yellow arrows == external factors.}
   \label{fig:conceptualFramework}
    \vspace{-15pt}
\end{figure}


The field of Artificial Intelligence (AI) has experienced several historic hype cycles where failed returns on investment led to AI Winters~\cite{haugeland97}. 
During the expert systems boom of the 1980s, global investments exceeded \$2.5 billion~\cite{roland2002strategic, feigenbaum1993japanese, newquist1994brain}, but when these systems failed to deliver the expected returns, funding was dramatically cut, precipitating the second AI Winter~\cite{roland2002strategic,toosi2021brief}.
Today, AI experts and some economists argue that the transformational promise of Generative AI (GenAI) will create significant lasting impact across industry sectors~\cite{jovanovic22, chui23}.
GenAI's ability to augment processes that were previously difficult to digitize has lead to widespread optimism and the rapid deployment of GenAI tooling by many organizations, often with minimal strategic planning and high expectations regarding productivity gains and subsequent cost savings~\cite{mayer25,uplevel24,microsoft24,li24b}. 

Like many others, the software engineering (SE) domain has been profoundly impacted by the introduction of GenAI tooling~\cite{sichkarenko24,russo24,daigle24}.
\gait in SE show great potential, with some projections estimating productivity increases between 20\% to 55\%~\cite{chui23,peng23,kalliamvakou24}.
Yet despite these projections, studies report strikingly mixed outcomes in real-world usage~\cite{liang24,becker2025measuring,harding2024coding}. 
While some developers achieve the promised productivity gains~\cite{sergeyuk2025human}, others experience decreased efficiency when attempting to integrate these tools~\cite{butler25}.

In the face of these challenges, a rapidly-evolving body of research focused on the adoption, usage, and integration of \gaidt has emerged in the past several years cataloging both enablers and barriers~\cite{kuhail2024will,boucher2024resistance,liang24,banh25}.
Much of this work relies on individual-level models such as the Technology Acceptance Model (TAM)~\cite{davis1989technology} and Unified Theory of Acceptance and Use of Technology (UTAUT)~\cite{venkatesh2003user}. While such lenses surface beliefs like perceived usefulness, they rarely hold constant the external realities - such as codebase complexity and management - that strongly shape tool adoption ~\cite{butler25, lambiase25, fulk93}.
With the goal of understanding what differentiates developers who are frequent versus infrequent \gaidts users within the same team contexts, we begin by exploring the research question (RQ):

\begin{itemize}
     \item[{\small \textbf{\rqone}}] What individual factors distinguish frequent and infrequent users of Generative AI development tools?
 \end{itemize}
Early discussions revealed that team and organizational factors appeared to shape individual factors for some developers (\cf Figure~\ref{fig:conceptualFramework}), leading us to explore these emergent results with a second RQ:
 \begin{itemize}
     \item[{\small \textbf{\rqtwo}}] How do team and organizational factors influence individual developer's Generative AI development tool usage?
 \end{itemize}

We utilize a paired interview design, conducting sequential semi-structured interviews with 54 developers representing 27 pairs from the same team matched on key factors—primary programming language, role, and seniority- but who exhibit contrasting usage patterns (one frequent user, one infrequent), identified using telemetry data from a large multinational software company. 
This matching strategy ensures that both developers share the same team-level context—codebase, manager, policies—so we can directly compare the context-specific blockers an infrequent user faces with the work‑arounds their frequent teammate has discovered. 
In doing so we trace how organizational conditions are interpreted and acted on rather than assuming purely individual causes, providing an empirically-grounded sociotechnical account of \gaits usage. 


We identified key differences in approaches frequent and infrequent users (\cf Figure~\ref{fig:conceptualFramework}): frequent users often conceptualized the tooling as a collaborative partner, adopted continuous and experimental integration approaches, and leveraged adaptive persistence when facing challenges, while infrequent users more often viewed it as a utility feature, maintained conservative integration patterns, and more often quickly abandoned tools when facing challenges.
Our analysis also identified key commonalities between the two groups: for many developers, organizational factors can actively shape these individual factors through an amplification effect, \eg 
team-specific demonstrations of applying \gait on common development tasks can transform developers' perception of tool usefulness.
Most critically, we identified a phenomenon which we refer to as the \ppp: increased productivity expectations from management without corresponding support often create a paradoxical effect, where developers lack the time necessary to develop the skills that would save time. 
This finding challenges the prevailing \gaidts deployment strategy across the software industry, which frames the challenge of figuring out how to use these tools to yield the expected productivity gains as the responsibility of individual developers~\cite{mayer25,uplevel24,microsoft24,li24b}. 
Like historical technological transformations in manufacturing and software engineering, we argue systematic productivity gains require systematic organizational change, not individual workflow optimization~\cite{deming18,brooks74}. 
The ``deploy and figure it out'' strategy where companies invest billions into \gait without investing in the corresponding support structures necessary to enable the systematic evolution of development workflows that would potentially yield a return on investment, are largely driven by competitive pressures and repeat past failures to recognize that transformative technologies demand transformative organizational approaches~\cite{deming18,haveman19,taylor04}.

In summary, this paper makes the following contributions: (1) a conceptual theoretical framework detailing the differences in individual factors between frequent and infrequent \gaits users; (2) a taxonomy of the ways different team and organizational factors can shape individual factors that influence usage; and (3) the concept of the \ppp and evidence-based strategies organizations can adopt to overcome it.

\vspace{-4mm}

\section{Background and Related Work}
\label{sec:RelatedWork}
Given the rapid permeation of these tools in development work, the factors that affect developers’ adoption and integration of \gait have become a focal topic in the SE research community.

\subsection{Individual Factors Affecting Adoption}

Research has explored why developers use \gait or not, often by adapting or extending classic technology acceptance theories (e.g., UTAUT \cite{venkatesh2003user, venkatesh2012consumer}). Early evidence suggests that traditional adoption factors do not fully apply in developer-\gaits interaction contexts. For example, \citet{russo24} found that \gaits compatibility with a developers' workflow drives early adoption, outweighing traditional UTAUT factors such as effort expectancy.

Trust has emerged as a critical factor in the design and adoption of these tools \cite{wang2024investigating, johnson2023make}. 
\citet{choudhuri2024guides} found that system quality, functional value, and goal alignment increase developers' trust in GenAI tools, which—combined with risk tolerance, self-efficacy, and intrinsic motivations significantly influences adoption.

Additionally, empirical studies highlight the significant challenges developers face with current GenAI tooling support including inconsistent performance, poor code/output quality, suggestions that fail to meet contextual requirements, alongside interaction friction and sensemaking difficulties~\cite{choudhuri2025needs, liang24, choudhuri2024far, kuhail2024will}. 
These issues force developers into a cycle of constant re-prompting and result verification, placing a heavy burden on the user~\cite{choudhuri2025needs, butler25}. 
Developers must often invest high cognitive effort into crafting prompts and frequently edit or debug AI-generated code; actions which disrupt their ``flow'' state and can negate potential productivity gains~\cite{becker2025measuring,tang2023empirical,sergeyuk2025human}. 
They also raise ethical and legal concerns, including uncertainty about code provenance or licensing for AI-generated code, as well as concerns about biases or security vulnerabilities introduced by AI contributions~\cite{boucher2024resistance, banh25}.

Task differences can further complicate adoption~\cite{huy2024generative}. 
\citet{lambiase25} found that adoption drivers are not one-size-fits-all; they vary by task and can even negatively impact adoption in certain cases. For instance, peer opinions correlate with higher \gaits usage for decision-making tasks. Yet, strong social influence can backfire: developers might develop high expectations that the tool then fails to meet~\cite{lambiase25}. 
Moreover, some developers fear that over-reliance on AI could erode their own skills or even threaten job security in the long run~\cite{kuhail2024will}.
Indeed, expectations play a crucial role in guiding appropriate tool usage. How developers conceptualize the AI's capabilities can influence how they integrate it in their work~\cite{hoffman2023measures, weisz2023toward}. Prior work indicates that fostering an accurate understanding of an AI’s capabilities (and limitations) improves user trust and helps foster appropriate usage~\cite{weisz2023toward, amershi2019guidelines}.

Overall, the literature paints a narrative in which developers adopt \gait when it demonstrates value, integrates smoothly into workflows, and earns user trust. However, existing studies largely treat developers as a relatively uniform group or focus on formative adoption intentions. This framing limits the understanding of the presence of substantial environmental confounds such as team culture, codebase complexity, managerial attitudes, tool availability, and organizational norms, among others, which often vary across developers, making it difficult to disentangle individual attitudes from environmental influences. As a result, it remains unclear whether observed adoption patterns stem from personal disposition or from differences in the surrounding environment.

Our study addresses this gap directly through its paired interview design. By interviewing matched pairs of developers; each working on the same team and project, and with access to the same GenAI tools; we
can directly compare the blockers an infrequent user faces with the work‑arounds their frequent teammate has discovered. 
This design allows us to isolate the individual-level mindsets, strategies, and responses that shape divergent usage patterns, while revealing the influences of team and organizational scaffolding. We move beyond the individualistic lens of prior work and offer a sociotechnical account of GenAI usage that is both empirically grounded and practically actionable.


\vspace{-4mm}

\subsection{Environmental Factors Affecting Adoption}

Beyond individual attitudes, the environment in which a developer works (their team and organization) plays a critical role in GenAI adoption~\cite{prasad2024towards, korzynski2024trust, kurup2022factors}.  Decades of socio-constructivist research have shown that technology uptake is not merely a function of individual utility but a deeply sociotechnical process shaped by social context, organizational norms, and collective sensemaking~\cite{jones2008giddens, orlikowski1999technologies, rogers2014diffusion}. Fulk’s theory of the social construction of technology, for instance, emphasizes how coworkers’ beliefs and organizational cues shape perceptions of technological value~\cite{fulk93}.

In the SE domain, emerging studies echo these insights. Peer influence and leadership support have been shown to shape developers’ trust in GenAI tools~\cite{cheng2024would, korzynski2024trust}. 
For example, while top-down mandates alone do not guarantee adoption, strong managerial support (e.g., providing resources and training) can significantly boost trust~\cite{korzynski2024trust}.
Similarly, \citet{cheng2024would} show how developers in online communities collectively interpret GenAI tooling through shared experiences, with success stories boosting confidence and widely reported failures fostering caution.

However, some recent empirical work has also found mixed results when examining these dynamics \cite{russo24, lambiase2024investigating}. 
For instance, \citet{russo24} found limited support for the influence of organizational stance (external encouragement), a result that contrasts with broader theoretical expectations in organizational communication~\cite{fulk93,fulk1990social,schmitz1991organizational}. 
One explanation is that these studies are able to capture static slices of environmental influence through individual attitudes (\eg survey item on management support), failing to reflect how multifaceted team dynamics and organizational support influence adoption. This limitation can lead to counterintuitive findings (\eg a lack of significant evidence for role of organizational stance) that conflict with broader theoretical expectations (e.g., Fulk’s work showing co-worker beliefs strongly guide technology perceptions~\cite{fulk93}). 
Also, many studies focus on aggregate trends or initial intentions~\cite{lambiase2024investigating, lambiase25}.

We address this gap by taking a comprehensive, in-situ approach. By interviewing matched pairs of developers who share the same team and context but have contrasting usage patterns,
we can directly ask frequent users how they overcame the context-specific barriers faced by their peers. 
In doing so, we extend beyond affirming that ``environment matters” to unpacking which aspects of the environment---organizational/peer support, learning resources, etc.--- make a difference to developers. 
This nuanced understanding helps explain the divergence in usage trajectories even among developers in similar contexts.
Identifying these differences can guide teams and organizations in creating environments that support sustained, effective \gaits use in software development.



\vspace{-4mm}

\section{Research Design}
\label{sec:researchDesign}

To understand what distinguishes frequent and infrequent \gaidt users within similar organizational contexts, we performed semi-structured interviews with 54 developers representing 27 pairs from different teams across a large multinational software company. 
Below, we detail ethical considerations, study design, analysis strategy, and limitations. 

\vspace{-3mm}
\subsection{Ethical Considerations}
\label{subsec:ethicalConsiderations}
We recognize that studying technology adoption in contexts with strong organizational pressure requires exceptional and up front care to protect participants' wellbeing and privacy.
Given the company’s culture of heavily promoting \gaits adoption, we implemented multiple safeguards to protect all participants.

\mysec{Telemetry Data and Privacy} Tool usage telemetry is collected on an opt-out basis, with developers able to disable collection at any time. In sensitive contexts, opting out is explicitly encouraged.

\mysec{Protecting Within-Team Anonymity} To ensure participants from small teams were not identifiable to colleagues (especially infrequent users), we did 
not disclose that we interviewed other team members. 
We further framed the study as exploring ``diverse developer experiences with GenAI tools."

\mysec{Ensuring Non-Judgmental Interview Environments} 
We aimed to create psychological safety for participants whose usage or beliefs might diverge from organizational expectations, by vetting all questions for neutrality in terms of tool usage, reviewed by our team and external researchers. 
The first author, who is not affiliated with any tooling teams, conducted all interviews and emphasized confidentiality and ensured a thorough anonymization process.

\subsection{Identifying and Recruiting Participants}
\label{subsec:identifyingAndRecruitingParticipants}

Our study employed a paired interview design, recruiting developers from the same teams who exhibited contrasting usage patterns. This approach minimized the influence of team-specific factors—codebase complexity, development practices, and organizational culture—that have confounded previous studies~\cite{butler25, russo24, khojah24,lambiase25, liang24,banh25}. By comparing developers in closely similar environments, we isolated individual-level factors affecting usage patterns.


\mysec{Identifying Pairs of Developers}
To identify developer pairs with contrasting usage patterns but similar environments across the company, we extracted data about developers (that had not opted out) from internal telemetry databases containing demographic attributes and tool usage.
Across the company, Github Copilot is the officially sanctioned and freely available \gaidts~\cite{gitHubCopilot}.
The telemetry usage data for each developer indicates the number of days 
each developer used GitHub copilot during an eight-week observation window (from April 1st to May 31st 2025). 
We also collected their job title, career stage, employee level, geographic area code (\ie country), team within organizational hierarchy (\ie their management chain from the CEO down to their direct manager), and primary programming language.

Using this data set, we implemented a multi round pairing algorithm using the Hungarian method for optimal assignment and \Code{scipy.optimize}~\cite{mills07,spicyOptimize}.
First, the algorithm grouped developers by career stage, job title, employee level, country, and direct manager-- creating groups with similar professional contexts.
Within each group, the algorithm split developers into two subgroups at the median usage level before applying the assignment algorithm to maximize usage difference between pairs.


Once the pool of pairs had been identified, we filtered it to ensure each pair shared relevant characteristics while exhibiting significant differences in usage. 
Specifically, We selected pairs that had same primary programming language and a usage gap in the top third across all pairs (measured by the difference in days of GenAI tool usage between the pair's frequent and infrequent user).

\begin{figure}[t]
    \centering
    \includegraphics[width=\linewidth]{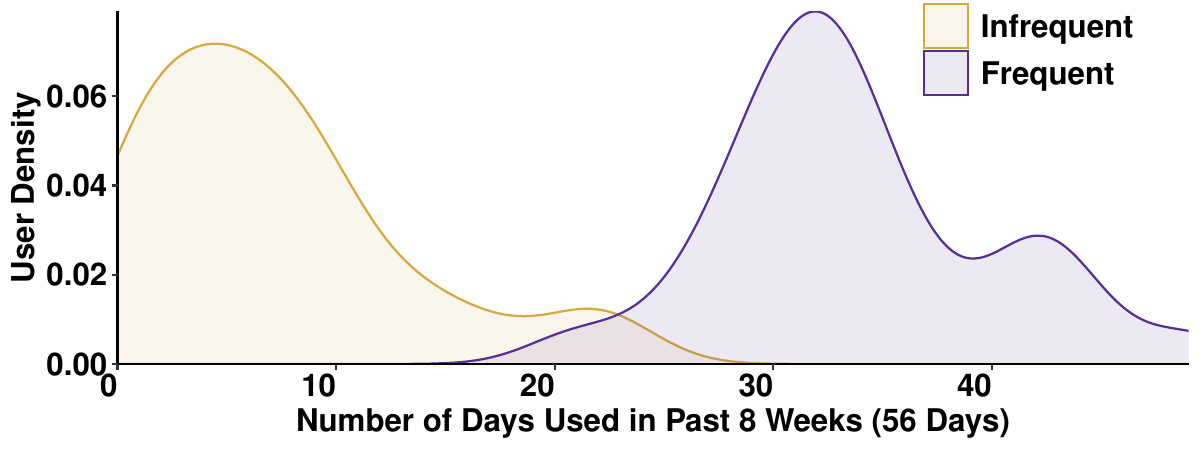}
 \caption{GenAI Tool Usage Distributions}
   \label{fig:participantUsageByGroup}
\end{figure}

Note our definitions of frequent and infrequent users are relative to their teams' usage (not to all developers across the company). 
We use the terms ``\lai'' and ``\hai'' to refer to the two groups of developers that represent the less frequent and more frequent users from each pair respectively.   
The median usage frequency for the infrequent and frequent groups was 5.5 days and 33 days during our observation window (\cf Figure~\ref{fig:participantUsageByGroup}).

\mysec{Participant Recruitment}
We recruited candidate pairs from engineering teams globally through an internal company chat, due to regulatory reasons we excluded pairs from Germany and Norway. 
We reached out to each developer individually, only interviewing developers from pairs where both agreed to participate.

\subsection{Interview Protocol}
\label{subsec:interviewProtocol}

Our interview protocol strategically leveraged the paired design. For each pair, we first interviewed the infrequent user to understand their perspectives, any barriers, and specific challenges they face. We then interviewed the corresponding frequent user from the same team, exploring how they may have navigated similar challenges and their broader experiences. This sequencing allowed us to probe the frequent users about context-specific barriers their teammates faced, revealing divergent responses and strategies they may have followed to overcome shared obstacles. The interview protocol evolved iteratively while maintaining consistency in core questions to ensure comparability across participants.\footnote{The complete interview guide is available in the supplementary material.} 


Our interview protocol initially drew from the UTAUT framework~\cite{venkatesh2003user}, a well-established lens in prior technology adoption research~\cite{venkatesh2003user, lambiase25,russo24}. 
UTAUT's four core constructs—performance expectancy, effort expectancy, social influence, and facilitating conditions—provided a systematic foundation for understanding usage behaviors in organizational contexts. 


We began with 30-minute interviews exploring these constructs through questions on perceived tool benefits (e.g., ``What are your current expectations about how GitHub Copilot can help with your development work?''), ease-of-use and integration challenges, alongside team influences, and organizational support. However, early interviews revealed patterns beyond traditional adoption factors— differences in how the two groups conceptualized and approached AI tool usage. To explore these emergent themes, we extended interviews to 60 minutes and supplemented UTAUT questions with new probes about participants' tool perceptions, integration strategies, learning approaches, and evolving skill perceptions.
\vspace{-2mm}
\subsection{Data Collection and Analysis}
\label{subsec:dataCollectionAndAnalysis}

\subfile{tables/participantDemographicTable.tex}

The interviews took place over video conferencing. 
From 670 invitations, 315 developers agreed to participate. Of those that agreed, 152 were from 76 pairs where both members agreed.  
We scheduled interviews with pairs on a first-come, first-served basis once both developers had agreed to participate, but were limited by protocol scheduling logistics, busy developer schedules, and time zones. 
In total, we conducted 56 interviews (labeled PID1-56), with 54 participants (27 matched pairs) included in the final analysis. 
Two participants were excluded when their pair partners dropped out.
We summarize the participant demographics in Table~\ref{tab:participantDemographics}.


We qualitatively analyzed interview transcript data using iterative thematic analysis~\cite{braun06}. 
Our process was guided by Lincoln and Guba's trustworthiness criteria~\cite{guba79}, as described by Nowell et al~\cite{nowell17}. 
During this process, we followed a commonly recommended strategy~\cite{lewis15}: 
throughout the process we switched between the different stages of 
analysis -- jumping between exploring the rich transcripts, 
engaging with and analytically memoing the data~\cite{miles14}, 
coding, searching for themes, and refining the codes and coding framework.
While UTAUT guided our interview questions and early memoing to organize emerging barriers, our actual coding framework was developed through iterative inductive analysis.\footnote{The complete coding framework and manual are available in the supplementary material.}

To leverage contextual richness, we coded both interviews from each pair in the same session, following the interview sequence (infrequent user first, then the frequent user). This paired approach allowed us to track how developers from closely similar team contexts responded differently to similar challenges. 
The analysis began with the first author performing open-ended inductive coding 
of each interview as data collection progressed. 
After eight interviews, all authors came together and 
performed an in-depth analysis of the codes and coding framework. 
Iterative adjustments to the coding framework and interview guide were 
made as necessary. 
Once the framework stabilized, the first author re-coded all transcripts, 
with uncertain cases being reviewed with another author's assistance.
We used HeyMarvin for qualitative analysis, 
facilitating code organization and pattern evolution tracking~\cite{heyMarvin}.
We ended the interviewing process when we reached our saturation criterion~\cite{francis10}, 
which we defined as two consecutive interview pairs without learning any 
new major insights (\ie 4 interviews).
For higher-level theme identification, we shifted focus from individual pairs to patterns distinguishing the two broader groups—frequent and infrequent users. This analytical pivot allowed us to identify systematic differences in how developers perceive, approach, and respond to AI tools despite shared organizational contexts, revealing individual factors that transcend team-specific circumstances. We report qualitative findings organized into emergent categories. We avoid discussing frequency counts or percentages in line with established guidance against quantifying qualitative data \cite{denzin2011sage}. 

\vspace{-2mm}

\subsection{Member Checking}
Following data analysis, we validated our findings through member checking with interviewees to ensure our interpretations accurately reflected their perspectives and experiences \cite{corbin14}. We sent all interviewees a draft of the paper's introduction, methods, results and discussion along with a list of questions asking for their extent of agreement with and/or the relevance of the findings, alongside additional thoughts and feedback.
Participant feedback confirmed our findings, with some providing additional clarifications, but no new insights or disagreements emerged \eg \participantQuote{I think it accurately reflects my experiences}{PID34}

\vspace{-3mm}

\subsection{Limitations}
\label{subsec:limitations}

Our interview study is affected by several limitations commonly experienced in interview research. The transferability of our findings may be influenced by self-selection bias among participants~\cite{rogelberg03, marcus05}, as there could be differences between those invited and those who participated. 
All participants were engineers at a large software company vocally committed to AI adoption, with GitHub Copilot as the only officially sanctioned tool. This limits generalizability to engineers at smaller or larger companies without AI promotion, or to open source development contexts. 
Despite this limitation, we argue our findings are still of value to the broader community because historically, 
it has been shown that single-case case studies can provide meaningful contributions to scientific discovery~\cite{flyvbjerg2006five}, 
and are an essential type of research in addition to research that studies broader populations as championed by Basili~\cite{basili02}.
A limitation to construct validity is that we measured AI tool usage through GitHub Copilot invocations only. Engineers using general-purpose AI tools (e.g., internal OpenAI models) or non-compliant tools (e.g., Cursor, Windsurf) could be classified as infrequent users despite potentially heavy AI usage elsewhere. As discussed in Section~\ref{subsec:ethicalConsiderations}, the company's strong pro-AI culture may have influenced participants' responses.
Finally, throughout this discussion, we occasionally use causal language (\eg `leads to') for clarity and readability. These phrasings should not be interpreted as statistical causal claims, as our qualitative design does not support formal causal inference.

\section{RQ1: What Distinguishes Frequent and Infrequent GenAI Development Tool Users}
\label{sec:RQ1Results}

\begin{figure}[t]
    \centering
    \includegraphics[width=.9\linewidth]{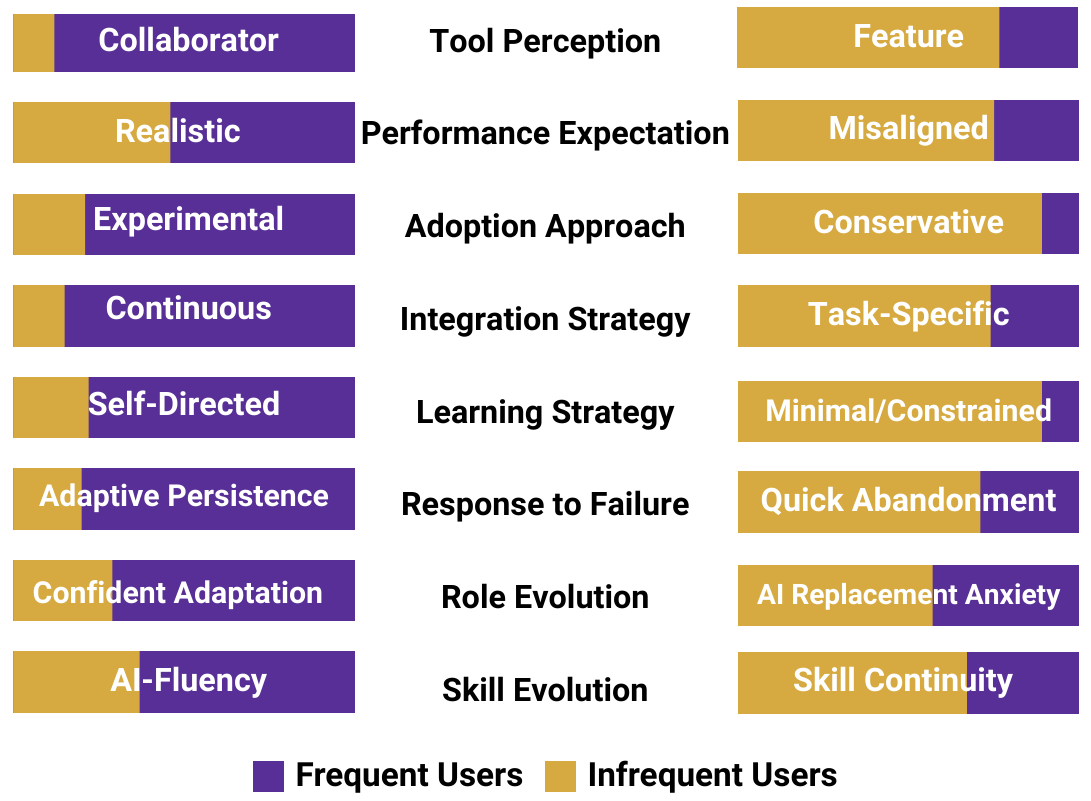}
   \vspace{-7pt}
 \caption{Distributions of user type proportions by factor.}
   \label{fig:codeFrequencyComparison}
    \vspace{-15pt}
\end{figure}

During our analysis, we identified distinct patterns in how frequent and infrequent users perceive, approach working with, and respond to challenges using \gaidt-- even among developers in similar organizational contexts (\cf Figure~\ref{fig:codeFrequencyComparison}). 
Figure~\ref{fig:conceptualFramework} illustrates our conceptual framework showing how individual tool integration progresses through distinct stages, with external factors influencing this process throughout (note that while arrows connect to specific stages for visual clarity, external factors impact the entire adoption journey).

\subsection{Mindset Formation: Initial Perceptions Shape Usage Patterns}
We observed two primary diverging patterns in tool perception that appeared to influence subsequent usage behaviors.

\subsubsection{Tool Perception: Collaborator vs. Feature}
\label{subsec:mindsetFormation}
Many \hai described their interactions with the tooling using collaborative language and partnership framing, using terms like ``teammate,'' ``collaborator,'' or ``assistant.''
\eg \participantQuote{I use GitHub as a colleague to discuss 'OK, this is what I'm trying to look at, this is the high level task, these are the my plans to go about implementing this feature, Do you see any gaps in that?'}{PID2}
This tool as a collaborator perception appeared to encourage deeper engagement as they invested time understanding its capabilities and strengths, as they would when onboarding a new team member.
\laic more often described these tools as enhanced features such as advanced code completion or search functionality \eg \participantQuote{I'm not going to say that I find it completely without value, but it's about as useful to me as a stack overflow search.}{PID7}
This utility framing appeared to limit engagement patterns— these developers tended to approach the tool with transactional expectations rather than collaborative ones.
The limiting nature of the utility framing is evident when comparing PID7 with their teammate PID23, who despite facing similar technical challenges, developed a more engaged usage pattern \eg \participantQuote{we have to write a lot of thorough tests that we can run on our gates periodically. So setting up those tests, writing the basic framework for them is just a lot of boilerplate code that GitHub Copilot helps me come up with... So that's one of the places I use it.}{PID23}


\subsubsection{Performance Expectations: Realistic vs. Misaligned}

Perception also appeared to influence performance expectations. 
\haic who viewed the tool as a collaborator often expressed realistic expectations, expecting assistance rather than replacement, similar to a junior colleague who provides value yet requires oversight: \participantQuote{I just have to type what I'm trying to do and then it gets me like 90\% of the way there and I have to just change a couple things.}{PID16}
\laic who viewed it as a feature often held misaligned expectations, either anticipating minimal value or expecting near-perfect functionality: \participantQuote{If I write a function, I would expect it to be able to write a perfect unit test for it.}{PID48} Both types of misaligned expectations appeared to lead to limited exploration or early abandonment when initial attempts failed.

\begin{keyInsightsBox}
{\textbf{Key Insights:}}
Frequent-AI users view GenAI as a collaborator. Infrequent users see it as a more or less useful developer tool.  
\end{keyInsightsBox}

\subsection{Approach Determination: How Developers Choose to Engage}
\label{subsec:adoptionApproach}
Following initial mindset formation, developers exhibited different interconnected approaches to tool usage (Figure~\ref{fig:conceptualFramework}), with experimental mindsets seemingly leading to broader usage patterns and conservative mindsets to isolated task-specific patterns.


\subsubsection{Usage Approach: Experimental vs. Conservative Patterns}
We observed two distinct approaches that shaped usage depth. 
Often \hai took an experimental approach, proactively trying new features, pushing tool boundaries, and experimenting to identify effective use cases \eg
\participantQuote{at first actually I just give every [one] of my tasks to the to the agent mode and ask him to write the code.}{PID55}
This experimental approach by PID55 contrasts with their teammate PID54's more conservative usage pattern, illustrating how risk tolerance and experimentation vary even among developers working on the same codebase
\haic tended to treat the tool as having undefined potential, actively seeking new applications and unconventional use cases.
\participantQuote{I started using images as well for doing CSS. Like, I'll open Paint and draw the layout I want, screenshot it, and feed that to Copilot to generate the CSS. It's very good.}{PID38}
This experimental approach led to continual usage with tools becoming deeply embedded in their daily workflows, reflecting a view of the tool as a general-purpose collaborator. 
The willingness to experiment created a positive feedback loop-- with successful use cases encouraging the exploration of others.

Alternatively, \lai were more likely to take a conservative approach, limiting tool usage to conventional well-vetted applications \eg \participantQuote{Something I've heard from a lot of people who've been using it is the first thing they tried out is writing tests using GitHub Copilot and I think that's probably the first thing that I actually tried out.}{PID8} 
A conservative approach appeared to couple naturally with task-specific usage limited to particular well-defined use cases, \eg 
\participantQuote{It is potentially useful for generating what I would describe as boilerplate code, and I find it frankly dangerous to use for anything beyond that.}{PID7}


\begin{keyInsightsBox}
{\textbf{Key Insights:}}
Tool perception shapes use: collaborative framing by \hai drives exploration, utility framing by \lai limits use to specific development tasks.
\end{keyInsightsBox}

\subsection{Encounters with Challenges: Universal Experiences, Divergent Responses}
\label{subsec:responseToFailure}
All participants reported encountering challenges with tool usability, particularly around prompt engineering and identifying effective use cases. The divergence emerged not in whether they faced difficulties, but rather in how they interpreted them and responded. 

\subsubsection{Response to Failure: Adaptive Persistence vs. Quick Abandonment}
Many \hai demonstrated adaptive persistence when facing limitations, continuing with modified approaches and treating failures as solvable puzzles requiring refinement strategies—such as breaking tasks into simpler subtasks or leveraging AI to improve their prompts \eg \participantQuote{If I give a big enough task to Copilot, it would hallucinate a lot. So it often needs to be broken down into smaller components. And once I do that, at times it does a pretty decent job.}{PID27}
In the face of poor tool performance for a particular task, they were also more likely to strategically select tasks for tooling assistance based on the tool's strengths and weaknesses, rather than abandoning or writing off the tooling entirely \eg 
\participantQuote{It's more just working with it, not exactly that I found a way to optimize it or make it better, but more like, `OK, these are the limitations, how do I work with it?'... These are some tasks that it can do. So OK, let me use it for those specific purposes.}{PID2}

Many \lai showed different response patterns to identical challenges. 
Compared to \hai who rarely exhibited quick abandonment when facing a challenge, \lai were much more likely to do so (\cf Figure~\ref{fig:codeFrequencyComparison}) \eg \participantQuote{I did something recently with the agent... It involved changing a lot of files in a similar manner. And it did the the change, let's say changing 15 files, but I had some compile errors and then I gave up on this approach and I took the slow and methodic way.}{PID24}
This quick abandonment is particularly striking when contrasted with PID24's own teammate PID26, who faced similar prompt engineering challenges but evolved their approach: \participantQuote{At the beginning I was using more the Google style. Write a sentence or a few words about what I want to search but... when using GitHub Copilot you need to tell a short story. Here is what I want to do. Here is what I tried so far. I failed with this. What would be another option [to go] about it?}{PID26}
Particularly among \lai with pessimistic tool expectations, many appeared to interpret these failures as confirmation of tool limitations, rather than as learning opportunities. 


\subsubsection{Learning Strategies: Self-Directed vs. Constrained Engagement}
Patterns in response to failure appeared to directly connect to learning strategies. 
Many \hai were self-directed learners pursuing adaptive persistence who naturally sought multiple learning resources. \participantQuote{I spent time reading blog posts, watching YouTube videos, even looking at how other people structure their prompts on GitHub. There's actually a lot to learn about how to use these tools effectively.}{PID8}
This multi-modal approach—combining formal documentation, community resources, and peer examples—accelerated skill development.
As with any tool, the learning curve to proficiency takes time and a common pattern among \hai was the recognition that carving out the time to do so, even when impending deadlines made it difficult, was worthwhile.

Conversely, \lai were more likely to engaged in minimal or constrained learning often due to time, resource, or motivational constraints \eg \participantQuote{Exploration wise it's been minimal. It kind of just whatever is obvious from the chat window and then mainly just that or completion. So stuff that's kind of like very low friction.}{PID6}
Without dedicated learning efforts, some \lai remained stuck in limited usage patterns, reinforcing their perception of minimal tool value within their development workflows.

\begin{keyInsightsBox} 
{\textbf{Key Insights:}}
When challenged, \hai persist and engage in self-directed learning, while \lai disengage with AI tools more quickly during development tasks.
\end{keyInsightsBox}

\subsection{Integration and Evolution: Long-term Professional Impact}
\label{subsec:roleEvolution}
Developers exhibited different visions of GenAI's impact on their professional identity and skills, which appeared to shape their integration efforts and skill development priorities.

\subsubsection{Role Conceptualization: Evolution and Competitive Pressure}
Both frequent and infrequent users recognized GenAI's potential to transform the developer role, though their responses differed. 
For many \lai the impact of this paradigm shift manifested as generalized concerns regarding GenAI making certain developer roles obsolete: \participantQuote{I think it will take entry level jobs now [that] it has reached the level where you can assign GitHub Copilot a ticket on Azure DevOps and it can create a pull request for you itself.}{PID44} 
The introduction of GenAI has created uncertainty regarding career progression especially for some early-career developers: \participantQuote{I am a junior developer right now, what does that mean in three years, four years? I do not know. I don't know if it's still coding and being technically viable or is it a different skill? So I think it has introduced a very important conversation on yeah, maybe pivoting.}{PID25} 

For \hai, concerns regarding the disruptive potential of GenAI often manifested as motivation to stay as in-the-know as possible.
Many were motivated to improve their proficiency with the tooling so they could stay competitive with other developers-- whomst they viewed as the immediate threat \eg \participantQuote{At the end of the day everyone is using it and everyone is being more productive and you wanna be one of those [people] otherwise [you're] gonna be left behind.}{PID20}
\haic recognized the same threat as \lai but often channeled it into motivation for skill development. They often exhibited confident adaptation, believing GenAI would enhance rather than replace their role, with some actively reframing their mindset positively given the perceived inevitability of tool integration \participantQuote{Viewing it as an extension of myself and trying to make the most of that versus trying to not be replaced... I think using it that way has been encouraging because if I view it as just amplifying me, multiplying me, not replace me, that gives me more encouragement, more energy to pursue it freely and not having that fear in the back of my head or these negative thoughts.}{PID6}
This belief also appeared to suggest a full circle impact back onto tool perception as visualized in Figure~\ref{fig:conceptualFramework} \participantQuote{I would think of it as a helper to make work efficient and easier and faster. So I'll say the job description would be software developer helper or assistant.}{PID29}

\vspace{-2mm}

\subsubsection{Skill Evolution: AI-Fluency vs. Continuity}
The different responses to competitive pressure manifested in contrasting approaches to skill development. Most \hai recognized AI fluency as an essential new competency requiring active cultivation \eg \participantQuote{I think it's really important that you know how to prompt well and ask the right questions. I think that's gonna be very useful for the future.}{PID12}
They invested in developing prompt engineering abilities, context management skills, and judgment about when and how to leverage AI assistance.
\laic often prioritized traditional skills while acknowledging potential changes.

\begin{keyInsightsBox} 
{\textbf{Key Insights:}}
Developers see AI as transformative, but frequent users pursue fluency and advantage, while \lai focus on job security and preserving traditional skills.
\end{keyInsightsBox}

\section{RQ2: How Team and Organizational Factors Shape Individual GenAI Use by Developers}
\label{sec:RQ2Results}

While our paired interview design revealed why developers in closely similar contexts showed different usage patterns by exploring context-specific barriers, these external factors do not affect all team members uniformly. 
Emergent findings from our analysis revealed that team and organizational factors do not just support or constrain usage-- they appear to actively shape individual factors for some developers.
Yet in every pair we studied, one developer was a frequent tool user, the other was not.
Initially, we expected that developers within similar contexts might show more uniform responses to these shared conditions. 
Instead, this emerging finding illuminated something more nuanced: These shared organizational factors interact with the individual differences identified in RQ1 in complex, emergent ways.
Social construction theory demonstrates that individuals within the same work group systematically interpret organizational influences differently depending on factors such as personal characteristics and social positions \cite{fulk93, orlikowski1994technological}.
Our results mirror this: we see how similar team contexts produce divergent responses through differential sensemaking processes \cite{weick1990technology, orlikowski1994technological, venkatesh2003user}.
This sociotechnical dynamic manifested through several interconnected mechanisms that appear to influence developers' usage journeys from initial perception through long-term integration.

\vspace{-2mm}

\subsection{The Impact of Leadership Communication} 
\label{subsec:messagingImpact}
Clear encouragement and communication from leadership regarding tool usage and value often provided developers with the confidence, permission, and motivation to invest resources into learning and exploration.
As one \hais explained: \participantQuote{Encouragements at the org level itself to share examples of how AI is making you better at work I think that was like a big thing}{PID2}
When managers shared concrete examples and tips, it provided practical value that motivated exploration for some developers: \participantQuote{I mean, they've shared tips and tricks that they found useful, which honestly have been helpful at times.}{PID23}
The regularity and consistency of messaging also appeared to matter in some cases.
After stalling in their usage after experiencing challenges, one developer found renewed motivation through recent communications from management: \participantQuote{there's been more messaging in the last month to try it out so I'm going to get back into it and see if I can find a way of using it that actually makes me more productive}{PID48}
Conversely, the absence of clear messaging left some developers uncertain about whether to integrate AI tools into workflows and what the value add would be. Several \lai described environments where they received minimal guidance which appeared to reinforce limited usage patterns: \participantQuote{Just it's a tool, [if] we can use then use it. But [leadership] haven't really been pushing... So we haven't really done much with that}{PID37}

\begin{keyInsightsBox} 
{\textbf{Key Insights:}}
Consistent leadership messaging encourages AI exploration. Without it, some developers are hesitant to use AI.
\end{keyInsightsBox} 

\subsection{Providing Context-Specific Resources}
\label{subsec:context-specificResources}
Many developers struggled with filtering through the large volume of ever-changing learning resources available to identify relevant resources worth investing their limited time into \eg \participantQuote{Some weeks I have 10+ meetings show up on my calendar about AI learning from all sorts of different places. And I can't go to all 10 so it becomes almost a little overwhelming where you just don't want to do any of them.}{PID33}
While some found resources on general topics like the basics of prompt engineering helpful, especially in early learning stages, such resources often failed to help support many developers through the transition between tool experimentation and meaningful effective integration into their workflows in practice. 
Many frequent and infrequent users expressed the need for context-specific guidance, calling for team-specific resources identifying suitable tasks and step-by-step implementation instructions, which one participant described as `\textit{AI-onboarding}' \eg \participantQuote{I mean like magic wand would be a comprehensive guide of how to use it for the specific projects that we work in with like example tasks that you could accomplish with it.}{PID48}


A lack of context-specific guidance created uncertainty about tool capabilities and appropriate use cases even among highly-motivated developers \eg \participantQuote{everyone wants to try GitHub Copilot, but right now they're still confused [about] what Copilot can do and how they can achieve their goals through [it].}{PID52}
Particularly when combined with strong encouragement to use the tool, a lack of concrete relevant guidance lead to frustration for some developers \eg \participantQuote{So this is where I get a bit jaded. A lot of the time what we hear from leadership is just `use AI, use AI, and AI is going to solve everything.'... I see a lot of talking about `we should use AI', I don't see a lot of useful collaboration.}{PID17}

Team-specific demonstrations were perceived as quite valuable and informative, with both participants from one team discussing a demonstration from a colleague who presented a shared-screen style step-by-step demonstration on how they used the tooling to accomplish a specific task commonly performed by the team \participantQuote{there's a very common pattern and structure that we have and he was showing how to do the prompts to use agent mode to make a brand new UX component and it saves a lot of time and so it was a very helpful demo.}{PID6}
The specificity of these demonstrations—showing exactly how to accomplish common team tasks—made the value immediately apparent and addressed the critical question of which tasks to use AI tools for and how.

\begin{keyInsightsBox} 
{\textbf{Key Insights:}}
Developers struggle to connect generic AI guidance to their work; team-specific demos show how AI may apply to their work.
\end{keyInsightsBox} 

\vspace{-2mm}

\subsection{Building Social Learning Structures} 
\label{subsec:socialLearningStructures}
Social learning structures helped create environments where individual discoveries benefited entire teams.
When teams had robust sharing practices, individual discoveries could create momentum for broader adoption.
Effective knowledge sharing took various forms, from formal sessions to informal exchanges, and these structures built collective competence and resilience in multiple ways.

Different structures served complimentary purposes. AI evangelists and champions helped some developers set realistic expectations about tool capabilities. By openly sharing both successes and limitations, they helped calibrate expectations appropriately.

Formal sessions provided structured learning and helped support skill development for some developers: \participantQuote{I would say [the] majority of my learning is from attending those sessions because most of the work we do is very similar in nature and it's quite applicable}{PID43}
When teams created spaces for shared learning, it transformed individual struggles into collective problem-solving opportunities.
However, peer influence operates differently even within teams. For example, PID48 found peer recommendations highly motivating: \participantQuote{Just hearing the recommendations from one of my co-workers is very motivating because if we're working as a good team, then we're all like having similar processes.}{PID48} 
Yet their own teammate PID35 remained an infrequent user despite this collaborative environment.

Informal knowledge sharing through daily interactions also shaped adoption journeys and enabled continuous micro-learning: \participantQuote{Sometimes you just get stuck, you go to somebody else and they're like, `Have you tried this feature?' And you're like, `I have not. Did it work for you?' And they're like, `Yes.' And that's how you end up trying [the feature], because somebody else tried it.}{PID18}
Micro-interactions accumulated into substantial support over time, creating an environment where exploration was socially reinforced.
However, developers without access to such structures faced different experiences. Some described learning in relative isolation: \participantQuote{I never share with anyone and I never received anything from my colleagues as well. I guess we all just kind of sit here and try to think that everyone else can use it.}{PID10}
Without contextual examples or peer support, the learning burden fell entirely on individuals \eg \participantQuote{I won't ask question on my team because I feel like using Copilot is more personal. It's not something my team can help me [with]}{PID56}

 \begin{keyInsightsBox} 
{\textbf{Key Insights:}}
Social learning structures can turn individual struggles into shared competencies; while knowledge isolation reinforces limited usage patterns.
\end{keyInsightsBox} 

\vspace{-3mm}

\subsection{Unexpected Interactions: Time, Expectations, and Developing Proficiency}
\label{subsec:unexpectedInteractions}

Sufficient learning time appeared crucial for both initial adoption and advanced skill development.
Many benefited from formal time allocations: \participantQuote{Our org has dedicated days to learn. We have around three to five days per semester where we can spend time on learning}{PID22} This legitimized exploration and removed guilt about skill development. 
Others leveraged scheduling flexibility to carve out their own time: \participantQuote{After seeing how powerful it is, I wanted to try it out myself. So one day I just thought OK... I will give one day [to] just focus on it. Try to learn what it is.}{PID21} 
Time during day-to-day work without deadline pressures proved essential for effective skill development and identifying effective use cases for many developers.
However, this individual initiative varies dramatically within teams. PID21 proactively carved out a full day for exploration, while their teammate PID8 took a more conservative approach, limiting exploration to use cases that others had validated.

A common theme among participants was experiencing increased productivity expectations from management because they have access to these tools \eg \participantQuote{I know that management is expecting us to produce code faster. And I know that the bar is gonna be raised every year because now, even though it's true or not true, [they] think that we should be much much more productive because we have these tools.}{PID20}
Some managers appeared to assume tool access automatically enabled rapid effective integration into existing workflows and immediate productivity gains, overlooking the significant learning effort and time investments required.
Many developers reported experiencing a circular challenge: \participantQuote{There's a skill gap among all of us about prompt crafting. The impression is that if I get better at prompt crafting then the Copilot could increase my productivity. But how do I get to that? How much productivity do I need to put into prompt crafting in order to increase productivity in my workflow? So there's a chicken or egg problem.}{PID46}

Increased productivity expectations which often translated into tighter deadlines prohibited many developers from doing the exploration necessary to learn how to use the tools effectively, instead falling back on their tried-and-true methods \eg \participantQuote{I do really feel the pressure of deadlines and trying to get stuff out and so feel less of the freedom to explore... The few times I've tried it, being discouraged, I'm like, OK, well, let me just get this done and then I'll get to it. That's been the hindrance.}{PID6}
Without time for exploration, many developers felt they couldn't understand the tool's full capabilities or how to use it most effectively: \participantQuote{we don't have so much time to learn about it because we are already overburdened with the tasks that we need to do}{PID44}

These contradictory experiences—heightened productivity expectations, significant learning time requirements, pressure-induced reversion to familiar methods, and limited support—formed a consistent pattern across many developers, suggesting systemic challenges in how some organizations approach GenAI tool integration.

\begin{keyInsightsBox} {\textbf{Key Insights:}} When organizations expect immediate gains from AI without learning support, the resulting pressure blocks the desired productivity benefits.   
\end{keyInsightsBox}

\section{Discussion and Implications}
\label{discussion}

\subsection{The Productivity Pressure Paradox}

\begin{figure}[t]
    \centering
    \includegraphics[height=4cm]{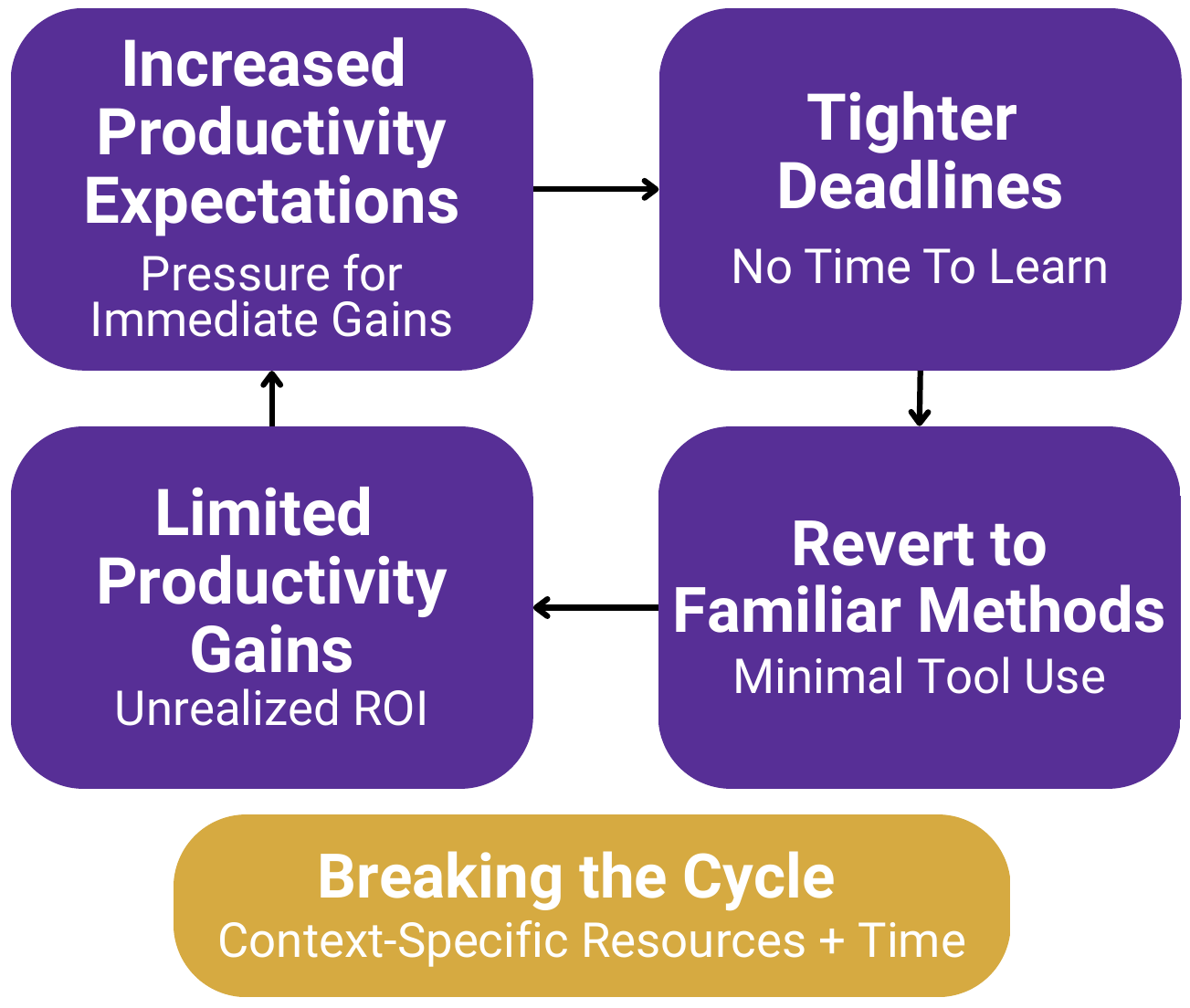}
   \vspace{-7pt}
 \caption{The Productivity Pressure Paradox.}
   \label{fig:pppFlowChart}
    \vspace{-15pt}
\end{figure}

Many organizations invest in \gaidt expecting productivity returns~\cite{chui23,apotheker24} while simultaneously treating adoption and integration into existing systems as an individual responsibility~\cite{mayer25,uplevel24,microsoft24,li24b}.
Yet within the context of knowledge work, the evolving capabilities of these tools create a ``jagged technological frontier,'' where they improve certain workflow tasks but can harm performance when applied to other seemingly comparable tasks~\cite{dell23}.
Many organizations are implicitly asking developers to navigate this jagged edge on their own while maintaining or increasing their productivity. 

This contradictory set of experiences—where developers must simultaneously maintain velocity, learn new tools, determine integration strategies, and navigate tool capability boundaries without systematic support—creates what we term the \ppp. 
This paradox creates a self-perpetuating organizational dynamic (\cf Figure~\ref{fig:pppFlowChart}): increased productivity expectations hinder the skill development investments needed to achieve those very gains through effective workflow integration, a challenge many participants faced (Section~\ref{subsec:unexpectedInteractions}).
Cognitive science reveals the underlying mechanism: time pressure fundamentally shifts individuals from flexible exploration to rigid exploitation patterns~\cite{wu22}, leading many developers to default to familiar approaches rather than discovering optimal integration strategies (Section~\ref{subsec:responseToFailure}).
The paradox appears to operate bidirectionally: first, productivity pressure prevents the temporary productivity dip necessary for learning; second, this approach places burden on individuals to solve organizational optimization challenges. 

\mysec{Increasing Productivity: From Systems to Individual Burdens}
The \ppp exemplifies how the knowledge sector uniquely places ``onto the individual worker the burden of improving output produced per unit of input''~\cite{newport21,perman10}-- a departure from established principles regarding productivity optimization which demonstrate that productivity improvements come from optimizing systems, not individual heroics~\cite{deming18,haveman19,taylor04}.
Henry Ford did not expect workers to individually discover how to build cars faster, he systematically redesigned the entire production process, investing significant time and effort into pioneering a reimagined system and building new tools to make it happen. 
The continuous-motion assembly line wasn't an incremental improvement from worker tinkering but a comprehensive system redesign that allowed individuals to focus on execution within an optimized workflow.
Workers were largely uninvolved in the optimization process; their productivity increased as a consequence of systemic change.

While the labor involved in pre-industrial automotive factories differs significantly from knowledge work like software engineering, this same principle is echoed in canonical software engineering literature. 
In the Mythical Man Month, Fred Brooks argues that adding developers to a late project does not speed it up but rather slows it down by adding layers of complexity~\cite{brooks74}.
To gain efficiency, better systems are needed. 
Brooks recognized that productivity improvements on a systematic level require systematic approaches rather than simply adding resources. 
The parallels become clearer when examining how many organizations approach GenAI integration today. Just as adding developers to a late project increases complexity without corresponding capability, adding GenAI without systematic support creates new coordination challenges and learning requirements (Section~\ref{subsec:responseToFailure}). 
Yet many organizations expect individual developers to navigate this complexity while maintaining delivery velocity, precisely the individualized approach to productivity that has proven ineffective across domains~\cite{lee21,harris94,ko19,cantrell23}.
We provide evidence for and echo the argument that the obligation to optimize productivity through GenAI integration should be shifted away from individuals and back toward systems. 

\subfile{tables/concreteReccomendationsTable.tex}

\vspace{-3mm}
\subsection{Implications for Organizations}
Our findings show how organizational factors can amplify individual usage patterns—reinforcing success when present or exacerbating difficulties when absent. 
These organizational approaches influence many developers' ability to navigate usage challenges.

Additionally, participant challenges align with key \textit{technostress} factors~\cite{brod84,vendramin21}. \textit{Techno-overload} manifested as increased productivity expectations regardless of actual proficiency. \textit{Techno-complexity} emerged through evolving tools creating steep learning curves. \textit{Techno-uncertainty} appeared as developers struggled to integrate tools into workflows without clear guidance.
Research demonstrates these factors harm productivity but can be mitigated through organizational support~\cite{fuglseth14,gaudioso17,vendramin21}. Task-technology fit theory shows that aligning technology capabilities, task requirements, and individual abilities requires systematic organizational intervention rather than individual adaptation alone~\cite{goodhue95}.


Organizations should reconceptualize how they assign responsibility for GenAI productivity optimization. 
The gap between what organizations expect and the support they provide can
create a counterproductive cycle, where the pressure for productivity gains risks undermining the desired results.
History offers a cautionary tale: the expert systems boom collapsed in part because organizations failed to provide adequate support infrastructure despite investing \$2.5 billion back in the 1980s, leading to the second AI winter when maintenance proved costlier than anticipated benefits~\cite{roland2002strategic,newquist1994brain,crevier1993ai}. 
Our findings echo this pattern, which suggests that organizations investing billions into GenAI technology might better realize returns by actively shaping systematic tool integration and investing in context-specific scaffolding that fosters usage and integration~\cite{toosi2021brief}.

\mysec{Increasing Productivity by Increasing Support}

Based on our findings, we identify interventions that helped participants navigate AI tool usage challenges. Table~\ref{tab:team-interventions} summarizes these evidence-based approaches.
Effective organizational messaging proved foundational for shaping tool perception and adoption. 
When tools were framed as collaborative `partners', developers often reported more successful engagement patterns.
Champions proved crucial, translating abstract capabilities into practical applications. These informal leaders created communities of practice where developers shared failures and successes without productivity judgment. \participantQuote{We share everything—the prompts that work, the ones that fail spectacularly, and most importantly, why}{PID47}

Context-specific resources resonated more than generic documentation. 
Developers benefited from domain-specific patterns, use case repositories tailored to their tech stack, and debugging strategies relevant to their codebase. 
Protected experimentation time—'AI Fridays' or temporarily suspended velocity expectations—enabled pressure free exploration without deadline pressures.
These interventions reinforce each other. The productivity pressure paradox emerges when support fails at any level, forcing individuals to bear organizational optimization burdens. 

\vspace{-3mm}
\subsection{Implications for Research}
Our findings raise important questions about temporal dynamics in GenAI adoption. While technology adoption is inherently dynamic~\cite{venkatesh2012consumer,rogers2014diffusion}, longitudinal research could reveal whether initial struggles give way to fluency as expertise theory suggests~\cite{ericsson1993role}, or if rapid AI innovation disrupts traditional learning curves. Does the productivity pressure paradox diminish as developers internalize usage patterns, or does it transform as tools evolve and expectations escalate? Research on technostress suggests pressures may compound rather than resolve~\cite{tarafdar2007impact}, indicating the paradox may evolve rather than disappear.

The mechanisms through which team factors shape individual approaches warrant deeper investigation. While team support amplifies adoption success or failure, specific mechanisms remain underexplored. Team influences alter individual behaviors through social learning~\cite{klein2000multilevel}, and organizational values cascade through levels~\cite{detert2000framework}, yet how this shapes AI tool usage remains unclear. Teams interpret mandates through local contexts~\cite{orlikowski2000using}—understanding these dynamics could inform more nuanced adoption strategies.

The distinction between viewing GenAI as "collaborator" versus "feature" reveals how mental models profoundly impact adoption trajectories. Research should explore how these models form through prior experiences, initial interactions, or social influences~\cite{gero2020mental}, and whether interventions like training or mentorship can shape productive mental models to prevent abandonment patterns~\cite{weisz2021perfection}.

Finally, our findings about shifting roles, from "code producers" to "AI orchestrators", highlight the need to rethink developer education. Successful AI adopters developed meta-cognitive skills: knowing when to use AI, crafting prompts, and critically evaluating outputs. Yet educational programs remain focused on traditional coding~\cite{becker2023programming}. How should we teach students to view AI as a partner rather than merely a tool~\cite{denny2024computing}? Research on these pedagogical shifts could prepare developers to navigate both the productivity pressure paradox and evolving development work.

\vspace{-3mm}
\section{Conclusion}
Our study challenges the prevailing narrative that genAI tool adoption is solely driven by individual initiative, revealing instead that successful adoption equally depends on the sociotechnical scaffolding provided by teams and organizations. Through interviews with 54 software developers, we show that developers thrive not just through personal mindset and learning strategies, but because their environments support experimentation, normalize failure, and offer recurring, context-specific guidance. We identify a \textit{Productivity Pressure Paradox}, where organizations expect gains from AI without investing in the necessary scaffolding infrastructure. Indeed, to realize genAI’s potential, we argue for a systemic shift in responsibility---from individuals to organizations; calling for environments that foster experimental learning through clear messaging and managerial support. Ultimately, the future of AI in software development will not be determined by tool capabilities alone, but by how well we design our support infrastructures around them.




\section{Data Availability}
\label{sec:dataAvailability}
Interview guide, codebook, and codebook manual are available in the supplemental materials on HotCRP. We will post the final supplementary materials publicly on Zenodo with the camera ready.

\begin{acks}

Heartfelt gratitude is bestowed upon Chanel~\includegraphics[height=1em]{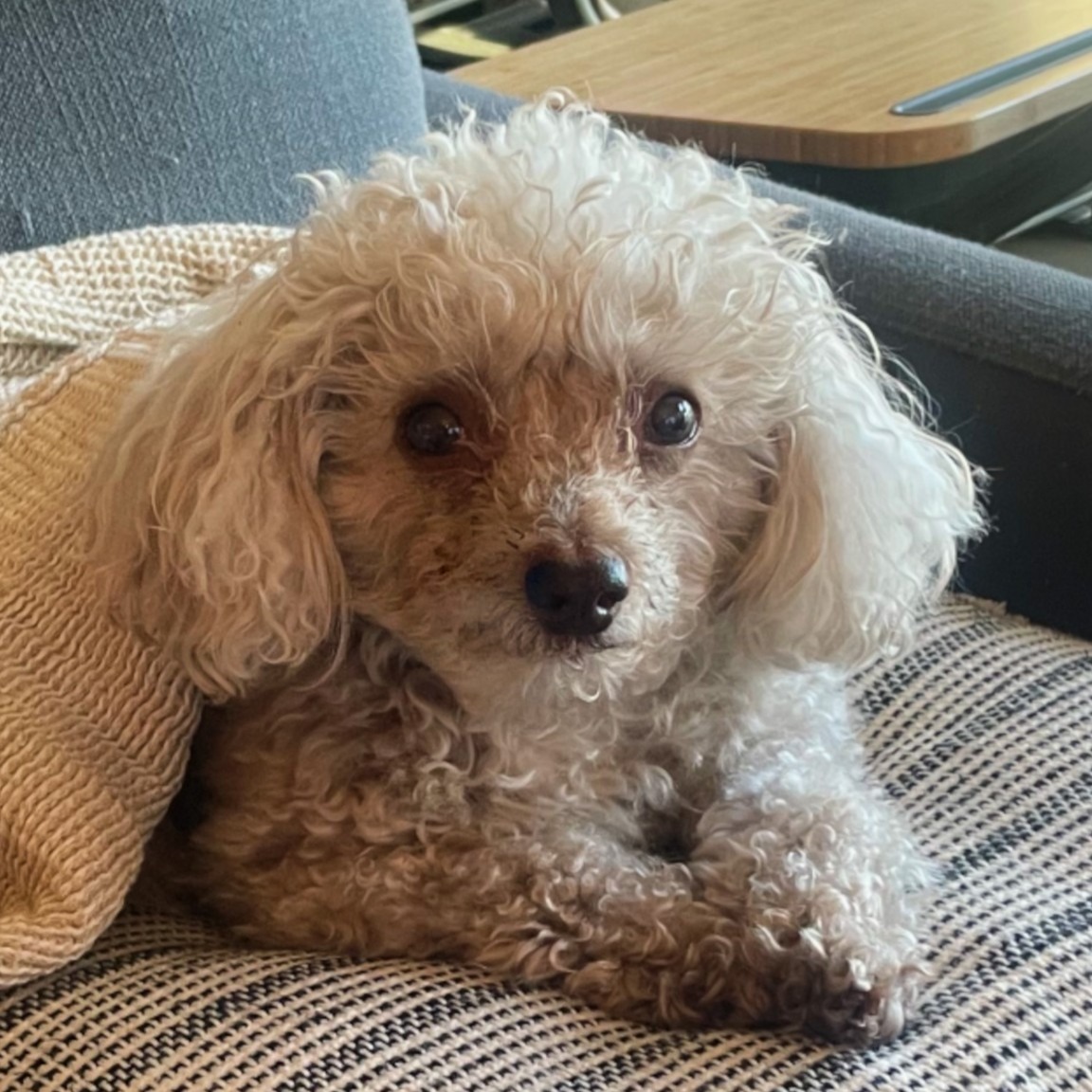} for braving a cross-country summer adventure in the name of continuing her ongoing work as a world-class canine researcher. 
Her brilliance, curiosity, and natural sense of wonder continue to propel the team toward progress.
Meet Pudding~\includegraphics[height=1em]{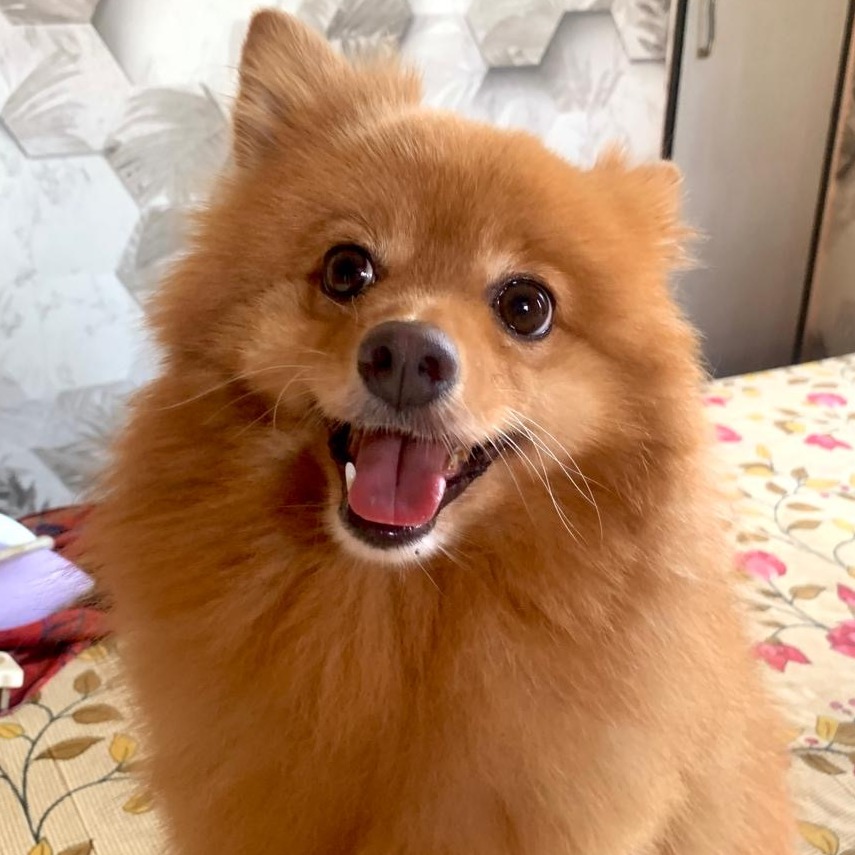}, the canine researcher. He has had an ever-enthusiastic presence during this project and is super proud of the work.
We thank the interview participants for generously sharing their expertise, perspective, and time with us, all insights in this work originated from them, and this work would not have been possible without them. 
Courtney Miller, Rudrajit Choudhuri, and Mara Ulloa performed this work during a summer internship at Microsoft. 
Margaret-Anne Storey contributed to this work while being a visiting researcher at Microsoft.
Special thanks are given to Brian Houck, Ben Hanrahan, and Travis Lowdermilk for providing valuable guidance on the interview protocol and exploring the emergent results.  
We thank Eirini Nathan for providing valuable insights and resources that helped shape the discussion section and the key message of the paper. 
We also thank Nancy Baym and Syboney Biwa for helping us contextualize our findings in the broader context of related fields of study and find valuable connections between them that helped strengthen this work. 
\end{acks}

\bibliographystyle{ACM-Reference-Format}
\bibliography{cmbib}


\end{document}

%% file: tables/participantDemographicTable.tex
\begin{table}[h]
\centering
\caption{Participant Demographics}
\vspace{-2mm}
\label{tab:participantDemographics}
\footnotesize
\begin{tabular*}{\columnwidth}{@{\extracolsep{\fill}}lclclc@{}}
\toprule
\multicolumn{2}{c}{\textbf{Location}} & \multicolumn{2}{c}{\textbf{Role}} & \multicolumn{2}{c}{\textbf{Copilot Use}} \\
\midrule
United States & 36 (66.7\%) & Senior SWE & 32 (59.3\%) & <6 months & 11 (20.4\%) \\
China & 6 (11.1\%) & SWE II & 18 (33.3\%) & 6-12 months & 9 (16.7\%) \\
Kenya & 4 (7.4\%) & SWE & 4 (7.4\%) & 1-2 years & 21 (38.9\%) \\
Canada & 2 (3.7\%) & & & 2-3 years & 13 (24.1\%) \\
Estonia & 2 (3.7\%) & & & & \\
Ireland & 2 (3.7\%) & & & & \\
Romania & 2 (3.7\%) & & & & \\
\bottomrule
\end{tabular*}
\end{table}

%% file: tables/concreteReccomendationsTable.tex
\begin{table*}[!ht]
\centering
\caption{Evidence-Based Team Interventions for AI Tool Adoption}
\vspace{-3mm}
\label{tab:team-interventions}
\footnotesize 
\renewcommand{\arraystretch}{1.3} 
\begin{tabularx}{\textwidth}{@{}>{\raggedright\arraybackslash}p{3.2cm}>{\raggedright\arraybackslash}p{2.8cm}>{\raggedright\arraybackslash}X>{\raggedright\arraybackslash}X@{}}
\toprule
\textbf{Intervention} & \textbf{Target Stage} & \textbf{Mechanism} & \textbf{Implementation Examples} \\
\midrule
\textbf{Clear Organizational Messaging} & 
Mindset Formation & 
Sets expectations; legitimizes exploration & 
Leadership demos; documented value propositions; success metrics \\
\addlinespace[2pt]

\textbf{AI Tool Champions} & 
Approach Determination & 
Models usage; provides guidance; reduces risk & 
1--2 designated team experts with 10\% time allocation for peer support \\
\addlinespace[2pt]

\textbf{Context-Specific Use Cases} & 
Experience \& Learning & 
Translates capabilities to team workflows & 
Team wiki with domain examples from actual repositories \\
\addlinespace[2pt]

\textbf{Knowledge Sharing Infrastructure} & 
Experience \& Learning & 
Accelerates learning; normalizes challenges & 
Dedicated channel; ``AI wins/fails'' in standups; pair programming \\
\addlinespace[2pt]

\textbf{Protected Learning Time} & 
Throughout & 
Enables pressure-free experimentation & 
``AI Fridays''; sprint time allocation; hackathons \\
\addlinespace[2pt]

\textbf{Collaborative Sessions} & 
Experience \& Learning & 
Transforms individual struggles to team learning & 
Weekly demos; monthly challenges; brown bags \\
\bottomrule
\end{tabularx}
\end{table*}

%% file: main.bbl

\begin{thebibliography}{96}


\ifx \showCODEN    \undefined \def \showCODEN     #1{\unskip}     \fi
\ifx \showISBNx    \undefined \def \showISBNx     #1{\unskip}     \fi
\ifx \showISBNxiii \undefined \def \showISBNxiii  #1{\unskip}     \fi
\ifx \showISSN     \undefined \def \showISSN      #1{\unskip}     \fi
\ifx \showLCCN     \undefined \def \showLCCN      #1{\unskip}     \fi
\ifx \shownote     \undefined \def \shownote      #1{#1}          \fi
\ifx \showarticletitle \undefined \def \showarticletitle #1{#1}   \fi
\ifx \showURL      \undefined \def \showURL       {\relax}        \fi
\providecommand\bibfield[2]{#2}
\providecommand\bibinfo[2]{#2}
\providecommand\natexlab[1]{#1}
\providecommand\showeprint[2][]{arXiv:#2}

\bibitem[git({[n.\,d.]})]%
        {gitHubCopilot}
 \bibinfo{year}{[n.\,d.]}\natexlab{}.
\newblock \bibinfo{title}{GitHub Copilot}.
\newblock \bibinfo{howpublished}{\url{https://github.com/features/copilot}}.
\newblock
\newblock
\shownote{Accessed Jun.\ 2025}.


\bibitem[hey({[n.\,d.]})]%
        {heyMarvin}
 \bibinfo{year}{[n.\,d.]}\natexlab{}.
\newblock \bibinfo{title}{HeyMarvin}.
\newblock \bibinfo{howpublished}{\url{https://heymarvin.com/}}.
\newblock
\newblock
\shownote{Accessed Jun.\ 2025}.


\bibitem[Amershi et~al\mbox{.}(2019)]%
        {amershi2019guidelines}
\bibfield{author}{\bibinfo{person}{Saleema Amershi}, \bibinfo{person}{Dan Weld}, \bibinfo{person}{Mihaela Vorvoreanu}, \bibinfo{person}{Adam Fourney}, \bibinfo{person}{Besmira Nushi}, \bibinfo{person}{Penny Collisson}, \bibinfo{person}{Jina Suh}, \bibinfo{person}{Shamsi Iqbal}, \bibinfo{person}{Paul~N Bennett}, \bibinfo{person}{Kori Inkpen}, {et~al\mbox{.}}} \bibinfo{year}{2019}\natexlab{}.
\newblock \showarticletitle{Guidelines for human-{AI} interaction}. In \bibinfo{booktitle}{\emph{2019 CHI Conference on Human Factors in Computing Systems}}. \bibinfo{pages}{1--13}.
\newblock


\bibitem[Apotheker et~al\mbox{.}({[n.\,d.]})]%
        {apotheker24}
\bibfield{author}{\bibinfo{person}{Jessica Apotheker} {et~al\mbox{.}}} \bibinfo{year}{[n.\,d.]}\natexlab{}.
\newblock \bibinfo{title}{From Potential to Profit with GenAIe}.
\newblock
\urldef\tempurl%
\url{https://www.bcg.com/publications/2024/from-potential-to-profit-with-genai}
\showURL{%
\tempurl}
\newblock
\shownote{Accessed: 2025-05-17}.


\bibitem[Banh et~al\mbox{.}(2025)]%
        {banh25}
\bibfield{author}{\bibinfo{person}{Leonardo Banh}, \bibinfo{person}{Florian Holldack}, {and} \bibinfo{person}{Gero Strobel}.} \bibinfo{year}{2025}\natexlab{}.
\newblock \showarticletitle{Copiloting the future: How generative AI transforms Software Engineering}.
\newblock \bibinfo{journal}{\emph{Information and Software Technology}}  \bibinfo{volume}{183} (\bibinfo{year}{2025}), \bibinfo{pages}{107751}.
\newblock


\bibitem[Basili et~al\mbox{.}(2002)]%
        {basili02}
\bibfield{author}{\bibinfo{person}{Victor~R Basili}, \bibinfo{person}{Forrest Shull}, {and} \bibinfo{person}{Filippo Lanubile}.} \bibinfo{year}{2002}\natexlab{}.
\newblock \showarticletitle{Building knowledge through families of experiments}.
\newblock \bibinfo{journal}{\emph{IEEE transactions on software engineering}} \bibinfo{volume}{25}, \bibinfo{number}{4} (\bibinfo{year}{2002}), \bibinfo{pages}{456--473}.
\newblock


\bibitem[Becker et~al\mbox{.}(2023)]%
        {becker2023programming}
\bibfield{author}{\bibinfo{person}{Brett~A Becker}, \bibinfo{person}{Paul Denny}, \bibinfo{person}{James Finnie-Ansley}, \bibinfo{person}{Andrew Luxton-Reilly}, \bibinfo{person}{James Prather}, {and} \bibinfo{person}{Eddie~Antonio Santos}.} \bibinfo{year}{2023}\natexlab{}.
\newblock \showarticletitle{Programming is hard-or at least it used to be: Educational opportunities and challenges of ai code generation}. In \bibinfo{booktitle}{\emph{Proceedings of the 54th ACM Technical Symposium on Computer Science Education V. 1}}. \bibinfo{pages}{500--506}.
\newblock


\bibitem[Becker et~al\mbox{.}(2025)]%
        {becker2025measuring}
\bibfield{author}{\bibinfo{person}{Joel Becker}, \bibinfo{person}{Nate Rush}, \bibinfo{person}{Elizabeth Barnes}, {and} \bibinfo{person}{David Rein}.} \bibinfo{year}{2025}\natexlab{}.
\newblock \showarticletitle{Measuring the Impact of Early-2025 AI on Experienced Open-Source Developer Productivity}.
\newblock \bibinfo{journal}{\emph{arXiv preprint arXiv:2507.09089}} (\bibinfo{year}{2025}).
\newblock


\bibitem[Boucher et~al\mbox{.}(2024)]%
        {boucher2024resistance}
\bibfield{author}{\bibinfo{person}{Josiah~D Boucher}, \bibinfo{person}{Gillian Smith}, {and} \bibinfo{person}{Yunus~Do{\u{g}}an Telliel}.} \bibinfo{year}{2024}\natexlab{}.
\newblock \showarticletitle{Is resistance futile?: Early career game developers, generative ai, and ethical skepticism}. In \bibinfo{booktitle}{\emph{Proceedings of the 2024 CHI Conference on Human Factors in Computing Systems}}. \bibinfo{pages}{1--13}.
\newblock


\bibitem[Braun and Clarke(2006)]%
        {braun06}
\bibfield{author}{\bibinfo{person}{Virginia Braun} {and} \bibinfo{person}{Victoria Clarke}.} \bibinfo{year}{2006}\natexlab{}.
\newblock \showarticletitle{Using thematic analysis in psychology}.
\newblock \bibinfo{journal}{\emph{Qualitative research in psychology}} \bibinfo{volume}{3}, \bibinfo{number}{2} (\bibinfo{year}{2006}), \bibinfo{pages}{77--101}.
\newblock


\bibitem[Brod(1984)]%
        {brod84}
\bibfield{author}{\bibinfo{person}{Craig Brod}.} \bibinfo{year}{1984}\natexlab{}.
\newblock \showarticletitle{Technostress: The human cost of the computer revolution}.
\newblock \bibinfo{journal}{\emph{(No Title)}} (\bibinfo{year}{1984}).
\newblock


\bibitem[Brooks(1974)]%
        {brooks74}
\bibfield{author}{\bibinfo{person}{Frederick~P Brooks}.} \bibinfo{year}{1974}\natexlab{}.
\newblock \showarticletitle{The mythical man-month}.
\newblock \bibinfo{journal}{\emph{Datamation}} \bibinfo{volume}{20}, \bibinfo{number}{12} (\bibinfo{year}{1974}), \bibinfo{pages}{44--52}.
\newblock


\bibitem[Butler et~al\mbox{.}(2025)]%
        {butler25}
\bibfield{author}{\bibinfo{person}{Jenna Butler}, \bibinfo{person}{Jina Suh}, \bibinfo{person}{Sankeerti Haniyur}, {and} \bibinfo{person}{Constance Hadley}.} \bibinfo{year}{2025}\natexlab{}.
\newblock \showarticletitle{Dear Diary: A randomized controlled trial of Generative AI coding tools in the workplace}. In \bibinfo{booktitle}{\emph{Proc.\ Int'l Conf.\ Software Engineering (ICSE)}}.
\newblock


\bibitem[Cantrell and Commisso({[n.\,d.]})]%
        {cantrell23}
\bibfield{author}{\bibinfo{person}{Sue Cantrell} {and} \bibinfo{person}{Corrie Commisso}.} \bibinfo{year}{[n.\,d.]}\natexlab{}.
\newblock \bibinfo{title}{Outcomes over outputs: Why productivity is no longer the metric that matters most}.
\newblock
\urldef\tempurl%
\url{https://www.deloitte.com/us/en/insights/topics/talent/measuring-productivity.html}
\showURL{%
\tempurl}


\bibitem[Cheng et~al\mbox{.}(2024)]%
        {cheng2024would}
\bibfield{author}{\bibinfo{person}{Ruijia Cheng}, \bibinfo{person}{Ruotong Wang}, \bibinfo{person}{Thomas Zimmermann}, {and} \bibinfo{person}{Denae Ford}.} \bibinfo{year}{2024}\natexlab{}.
\newblock \showarticletitle{“It would work for me too”: How online communities shape software developers’ trust in AI-powered code generation tools}.
\newblock \bibinfo{journal}{\emph{ACM Transactions on Interactive Intelligent Systems}} \bibinfo{volume}{14}, \bibinfo{number}{2} (\bibinfo{year}{2024}), \bibinfo{pages}{1--39}.
\newblock


\bibitem[Choudhuri et~al\mbox{.}(2024a)]%
        {choudhuri2024far}
\bibfield{author}{\bibinfo{person}{Rudrajit Choudhuri}, \bibinfo{person}{Dylan Liu}, \bibinfo{person}{Igor Steinmacher}, \bibinfo{person}{Marco Gerosa}, {and} \bibinfo{person}{Anita Sarma}.} \bibinfo{year}{2024}\natexlab{a}.
\newblock \showarticletitle{How far are we? the triumphs and trials of generative ai in learning software engineering}. In \bibinfo{booktitle}{\emph{Proceedings of the IEEE/ACM 46th international conference on software engineering}}. \bibinfo{pages}{1--13}.
\newblock


\bibitem[Choudhuri et~al\mbox{.}(2024b)]%
        {choudhuri2024guides}
\bibfield{author}{\bibinfo{person}{Rudrajit Choudhuri}, \bibinfo{person}{Bianca Trinkenreich}, \bibinfo{person}{Rahul Pandita}, \bibinfo{person}{Eirini Kalliamvakou}, \bibinfo{person}{Igor Steinmacher}, \bibinfo{person}{Marco Gerosa}, \bibinfo{person}{Christopher Sanchez}, {and} \bibinfo{person}{Anita Sarma}.} \bibinfo{year}{2024}\natexlab{b}.
\newblock \showarticletitle{What Guides Our Choices? Modeling Developers' Trust and Behavioral Intentions Towards GenAI}.
\newblock \bibinfo{journal}{\emph{arXiv preprint arXiv:2409.04099}} (\bibinfo{year}{2024}).
\newblock


\bibitem[Choudhuri et~al\mbox{.}(2025)]%
        {choudhuri2025needs}
\bibfield{author}{\bibinfo{person}{Rudrajit Choudhuri}, \bibinfo{person}{Bianca Trinkenreich}, \bibinfo{person}{Rahul Pandita}, \bibinfo{person}{Eirini Kalliamvakou}, \bibinfo{person}{Igor Steinmacher}, \bibinfo{person}{Marco Gerosa}, \bibinfo{person}{Christopher Sanchez}, {and} \bibinfo{person}{Anita Sarma}.} \bibinfo{year}{2025}\natexlab{}.
\newblock \showarticletitle{What Needs Attention? Prioritizing Drivers of Developers' Trust and Adoption of Generative AI}.
\newblock \bibinfo{journal}{\emph{arXiv preprint arXiv:2505.17418}} (\bibinfo{year}{2025}).
\newblock


\bibitem[Chui et~al\mbox{.}(2023)]%
        {chui23}
\bibfield{author}{\bibinfo{person}{Michael Chui}, \bibinfo{person}{Eric Hazan}, \bibinfo{person}{Roger Roberts}, \bibinfo{person}{Alex Singla}, {and} \bibinfo{person}{Kate Smaje}.} \bibinfo{year}{2023}\natexlab{}.
\newblock \showarticletitle{The economic potential of generative AI}.
\newblock  (\bibinfo{year}{2023}).
\newblock


\bibitem[Corbin and Strauss(2014)]%
        {corbin14}
\bibfield{author}{\bibinfo{person}{Juliet Corbin} {and} \bibinfo{person}{Anselm Strauss}.} \bibinfo{year}{2014}\natexlab{}.
\newblock \bibinfo{booktitle}{\emph{Basics of qualitative research: Techniques and procedures for developing grounded theory}}.
\newblock \bibinfo{publisher}{Sage publications}.
\newblock


\bibitem[Crevier(1993)]%
        {crevier1993ai}
\bibfield{author}{\bibinfo{person}{Daniel Crevier}.} \bibinfo{year}{1993}\natexlab{}.
\newblock \bibinfo{booktitle}{\emph{AI: the tumultuous history of the search for artificial intelligence}}.
\newblock \bibinfo{publisher}{Basic Books, Inc.}
\newblock


\bibitem[Daigle and Staff(2024)]%
        {daigle24}
\bibfield{author}{\bibinfo{person}{Kyle Daigle} {and} \bibinfo{person}{GitHub Staff}.} \bibinfo{year}{2024}\natexlab{}.
\newblock \bibinfo{booktitle}{\emph{Octoverse: The state of open source and rise of AI in 2023}}.
\newblock \bibinfo{type}{{T}echnical {R}eport}. \bibinfo{institution}{GitHub}.
\newblock
\urldef\tempurl%
\url{https://github.blog/news-insights/research/the-state-of-open-source-and-ai/}
\showURL{%
\tempurl}


\bibitem[Davis et~al\mbox{.}(1989)]%
        {davis1989technology}
\bibfield{author}{\bibinfo{person}{Fred~D Davis}, \bibinfo{person}{Richard~P Bagozzi}, {and} \bibinfo{person}{Paul~R Warshaw}.} \bibinfo{year}{1989}\natexlab{}.
\newblock \showarticletitle{Technology acceptance model}.
\newblock \bibinfo{journal}{\emph{J Manag Sci}} \bibinfo{volume}{35}, \bibinfo{number}{8} (\bibinfo{year}{1989}), \bibinfo{pages}{982--1003}.
\newblock


\bibitem[Dell'Acqua et~al\mbox{.}(2023)]%
        {dell23}
\bibfield{author}{\bibinfo{person}{Fabrizio Dell'Acqua}, \bibinfo{person}{Edward McFowland~III}, \bibinfo{person}{Ethan~R Mollick}, \bibinfo{person}{Hila Lifshitz-Assaf}, \bibinfo{person}{Katherine Kellogg}, \bibinfo{person}{Saran Rajendran}, \bibinfo{person}{Lisa Krayer}, \bibinfo{person}{Fran{\c{c}}ois Candelon}, {and} \bibinfo{person}{Karim~R Lakhani}.} \bibinfo{year}{2023}\natexlab{}.
\newblock \showarticletitle{Navigating the jagged technological frontier: Field experimental evidence of the effects of AI on knowledge worker productivity and quality}.
\newblock \bibinfo{journal}{\emph{Harvard Business School Technology \& Operations Mgt. Unit Working Paper}} \bibinfo{number}{24-013} (\bibinfo{year}{2023}).
\newblock


\bibitem[Deming(2018)]%
        {deming18}
\bibfield{author}{\bibinfo{person}{W~Edwards Deming}.} \bibinfo{year}{2018}\natexlab{}.
\newblock \bibinfo{booktitle}{\emph{Out of the Crisis, reissue}}.
\newblock \bibinfo{publisher}{MIT press}.
\newblock


\bibitem[Denny et~al\mbox{.}(2024)]%
        {denny2024computing}
\bibfield{author}{\bibinfo{person}{Paul Denny}, \bibinfo{person}{James Prather}, \bibinfo{person}{Brett~A Becker}, \bibinfo{person}{James Finnie-Ansley}, \bibinfo{person}{Arto Hellas}, \bibinfo{person}{Juho Leinonen}, \bibinfo{person}{Andrew Luxton-Reilly}, \bibinfo{person}{Brent~N Reeves}, \bibinfo{person}{Eddie~Antonio Santos}, {and} \bibinfo{person}{Sami Sarsa}.} \bibinfo{year}{2024}\natexlab{}.
\newblock \showarticletitle{Computing education in the era of generative AI}.
\newblock \bibinfo{journal}{\emph{Commun. ACM}} \bibinfo{volume}{67}, \bibinfo{number}{2} (\bibinfo{year}{2024}), \bibinfo{pages}{56--67}.
\newblock


\bibitem[Denzin and Lincoln(2011)]%
        {denzin2011sage}
\bibfield{author}{\bibinfo{person}{Norman~K Denzin} {and} \bibinfo{person}{Yvonna~S Lincoln}.} \bibinfo{year}{2011}\natexlab{}.
\newblock \bibinfo{booktitle}{\emph{The Sage handbook of qualitative research}}.
\newblock \bibinfo{publisher}{sage}.
\newblock


\bibitem[Detert et~al\mbox{.}(2000)]%
        {detert2000framework}
\bibfield{author}{\bibinfo{person}{James~R Detert}, \bibinfo{person}{Roger~G Schroeder}, {and} \bibinfo{person}{John~J Mauriel}.} \bibinfo{year}{2000}\natexlab{}.
\newblock \showarticletitle{A framework for linking culture and improvement initiatives in organizations}.
\newblock \bibinfo{journal}{\emph{Academy of management Review}} \bibinfo{volume}{25}, \bibinfo{number}{4} (\bibinfo{year}{2000}), \bibinfo{pages}{850--863}.
\newblock


\bibitem[Ericsson et~al\mbox{.}(1993)]%
        {ericsson1993role}
\bibfield{author}{\bibinfo{person}{K~Anders Ericsson}, \bibinfo{person}{Ralf~T Krampe}, {and} \bibinfo{person}{Clemens Tesch-R{\"o}mer}.} \bibinfo{year}{1993}\natexlab{}.
\newblock \showarticletitle{The role of deliberate practice in the acquisition of expert performance.}
\newblock \bibinfo{journal}{\emph{Psychological review}} \bibinfo{volume}{100}, \bibinfo{number}{3} (\bibinfo{year}{1993}), \bibinfo{pages}{363}.
\newblock


\bibitem[Feigenbaum and Shrobe(1993)]%
        {feigenbaum1993japanese}
\bibfield{author}{\bibinfo{person}{Edward Feigenbaum} {and} \bibinfo{person}{Howard Shrobe}.} \bibinfo{year}{1993}\natexlab{}.
\newblock \showarticletitle{The Japanese national Fifth Generation project: introduction, survey, and evaluation}.
\newblock \bibinfo{journal}{\emph{Future Generation Computer Systems}} \bibinfo{volume}{9}, \bibinfo{number}{2} (\bibinfo{year}{1993}), \bibinfo{pages}{105--117}.
\newblock


\bibitem[Flyvbjerg(2006)]%
        {flyvbjerg2006five}
\bibfield{author}{\bibinfo{person}{Bent Flyvbjerg}.} \bibinfo{year}{2006}\natexlab{}.
\newblock \showarticletitle{Five misunderstandings about case-study research}.
\newblock \bibinfo{journal}{\emph{Qualitative inquiry}} \bibinfo{volume}{12}, \bibinfo{number}{2} (\bibinfo{year}{2006}), \bibinfo{pages}{219--245}.
\newblock


\bibitem[Francis et~al\mbox{.}(2010)]%
        {francis10}
\bibfield{author}{\bibinfo{person}{Jill~J Francis} {et~al\mbox{.}}} \bibinfo{year}{2010}\natexlab{}.
\newblock \showarticletitle{What is an adequate sample size? Operationalising data saturation for theory-based interview studies}.
\newblock \bibinfo{journal}{\emph{Psychology and Health}} (\bibinfo{year}{2010}).
\newblock


\bibitem[Fuglseth and S{\o}reb{\o}(2014)]%
        {fuglseth14}
\bibfield{author}{\bibinfo{person}{Anna~Mette Fuglseth} {and} \bibinfo{person}{{\O}ystein S{\o}reb{\o}}.} \bibinfo{year}{2014}\natexlab{}.
\newblock \showarticletitle{The effects of technostress within the context of employee use of ICT}.
\newblock \bibinfo{journal}{\emph{Computers in human behavior}}  \bibinfo{volume}{40} (\bibinfo{year}{2014}), \bibinfo{pages}{161--170}.
\newblock


\bibitem[Fulk(1993)]%
        {fulk93}
\bibfield{author}{\bibinfo{person}{Janet Fulk}.} \bibinfo{year}{1993}\natexlab{}.
\newblock \showarticletitle{Social construction of communication technology}.
\newblock \bibinfo{journal}{\emph{Academy of Management journal}} \bibinfo{volume}{36}, \bibinfo{number}{5} (\bibinfo{year}{1993}), \bibinfo{pages}{921--950}.
\newblock


\bibitem[Fulk et~al\mbox{.}(1990)]%
        {fulk1990social}
\bibfield{author}{\bibinfo{person}{Janet Fulk}, \bibinfo{person}{Joseph Schmitz}, {and} \bibinfo{person}{Charles~W Steinfield}.} \bibinfo{year}{1990}\natexlab{}.
\newblock \showarticletitle{A social influence model of technology use}.
\newblock In \bibinfo{booktitle}{\emph{Organizations and communication technology}}. \bibinfo{publisher}{SAGE Publications, Inc.}, \bibinfo{pages}{117--140}.
\newblock


\bibitem[Gaudioso et~al\mbox{.}(2017)]%
        {gaudioso17}
\bibfield{author}{\bibinfo{person}{Fulvio Gaudioso}, \bibinfo{person}{Ofir Turel}, {and} \bibinfo{person}{Carlo Galimberti}.} \bibinfo{year}{2017}\natexlab{}.
\newblock \showarticletitle{The mediating roles of strain facets and coping strategies in translating techno-stressors into adverse job outcomes}.
\newblock \bibinfo{journal}{\emph{Computers in Human Behavior}}  \bibinfo{volume}{69} (\bibinfo{year}{2017}), \bibinfo{pages}{189--196}.
\newblock


\bibitem[Gero et~al\mbox{.}(2020)]%
        {gero2020mental}
\bibfield{author}{\bibinfo{person}{Katy~Ilonka Gero}, \bibinfo{person}{Zahra Ashktorab}, \bibinfo{person}{Casey Dugan}, \bibinfo{person}{Qian Pan}, \bibinfo{person}{James Johnson}, \bibinfo{person}{Werner Geyer}, \bibinfo{person}{Maria Ruiz}, \bibinfo{person}{Sarah Miller}, \bibinfo{person}{David~R Millen}, \bibinfo{person}{Murray Campbell}, {et~al\mbox{.}}} \bibinfo{year}{2020}\natexlab{}.
\newblock \showarticletitle{Mental models of AI agents in a cooperative game setting}. In \bibinfo{booktitle}{\emph{Proceedings of the 2020 chi conference on human factors in computing systems}}. \bibinfo{pages}{1--12}.
\newblock


\bibitem[Goodhue and Thompson(1995)]%
        {goodhue95}
\bibfield{author}{\bibinfo{person}{Dale~L Goodhue} {and} \bibinfo{person}{Ronald~L Thompson}.} \bibinfo{year}{1995}\natexlab{}.
\newblock \showarticletitle{Task-technology fit and individual performance}.
\newblock \bibinfo{journal}{\emph{MIS quarterly}} (\bibinfo{year}{1995}), \bibinfo{pages}{213--236}.
\newblock


\bibitem[Guba(1979)]%
        {guba79}
\bibfield{author}{\bibinfo{person}{Egon Guba}.} \bibinfo{year}{1979}\natexlab{}.
\newblock \showarticletitle{Naturalistic inquiry.}
\newblock \bibinfo{journal}{\emph{Improving Human Performance Qtrly.}} (\bibinfo{year}{1979}).
\newblock


\bibitem[Harding and Kloster(2024)]%
        {harding2024coding}
\bibfield{author}{\bibinfo{person}{William Harding} {and} \bibinfo{person}{Matthew Kloster}.} \bibinfo{year}{2024}\natexlab{}.
\newblock \showarticletitle{Coding on copilot: 2023 data suggests downward pressure on code quality}.
\newblock \bibinfo{journal}{\emph{https://www. gitclear. com/coding\_on\_copilot\_data\_shows\_ais\_downward\_pressure\_on\_code\_quality/}} (\bibinfo{year}{2024}).
\newblock


\bibitem[Harris(1994)]%
        {harris94}
\bibfield{author}{\bibinfo{person}{Douglas~H Harris}.} \bibinfo{year}{1994}\natexlab{}.
\newblock \bibinfo{booktitle}{\emph{Organizational linkages: Understanding the productivity paradox}}.
\newblock \bibinfo{publisher}{National Academies Press}.
\newblock


\bibitem[Haugeland(1997)]%
        {haugeland97}
\bibfield{author}{\bibinfo{person}{John Haugeland}.} \bibinfo{year}{1997}\natexlab{}.
\newblock \bibinfo{booktitle}{\emph{Mind Design II}}.
\newblock \bibinfo{publisher}{MIT Press}.
\newblock


\bibitem[Haveman and Wetts(2019)]%
        {haveman19}
\bibfield{author}{\bibinfo{person}{Heather~A Haveman} {and} \bibinfo{person}{Rachel Wetts}.} \bibinfo{year}{2019}\natexlab{}.
\newblock \showarticletitle{Organizational theory: From classical sociology to the 1970s}.
\newblock \bibinfo{journal}{\emph{Sociology Compass}} \bibinfo{volume}{13}, \bibinfo{number}{3} (\bibinfo{year}{2019}), \bibinfo{pages}{e12627}.
\newblock


\bibitem[Hoffman et~al\mbox{.}(2023)]%
        {hoffman2023measures}
\bibfield{author}{\bibinfo{person}{Robert~R Hoffman}, \bibinfo{person}{Shane~T Mueller}, \bibinfo{person}{Gary Klein}, {and} \bibinfo{person}{Jordan Litman}.} \bibinfo{year}{2023}\natexlab{}.
\newblock \showarticletitle{Measures for explainable {AI}: Explanation goodness, user satisfaction, mental models, curiosity, trust, and human-{AI} performance}.
\newblock \bibinfo{journal}{\emph{Frontiers in Computer Science}}  \bibinfo{volume}{5} (\bibinfo{year}{2023}), \bibinfo{pages}{1096257}.
\newblock


\bibitem[Huy et~al\mbox{.}(2024)]%
        {huy2024generative}
\bibfield{author}{\bibinfo{person}{Le~Van Huy}, \bibinfo{person}{Hien~TT Nguyen}, \bibinfo{person}{Tan Vo-Thanh}, \bibinfo{person}{Nguyen Huu~Thai Thinh}, \bibinfo{person}{Tran Thi Thu~Dung}, {et~al\mbox{.}}} \bibinfo{year}{2024}\natexlab{}.
\newblock \showarticletitle{Generative AI, why, how, and outcomes: A user adoption study}.
\newblock \bibinfo{journal}{\emph{AIS Transactions on Human-Computer Interaction}} \bibinfo{volume}{16}, \bibinfo{number}{1} (\bibinfo{year}{2024}), \bibinfo{pages}{1--27}.
\newblock


\bibitem[Johnson et~al\mbox{.}(2023)]%
        {johnson2023make}
\bibfield{author}{\bibinfo{person}{Brittany Johnson}, \bibinfo{person}{Christian Bird}, \bibinfo{person}{Denae Ford}, \bibinfo{person}{Nicole Forsgren}, {and} \bibinfo{person}{Thomas Zimmermann}.} \bibinfo{year}{2023}\natexlab{}.
\newblock \showarticletitle{Make your tools sparkle with trust: The PICSE framework for trust in software tools}. In \bibinfo{booktitle}{\emph{2023 IEEE/ACM 45th International Conference on Software Engineering: Software Engineering in Practice (ICSE-SEIP)}}. IEEE, \bibinfo{pages}{409--419}.
\newblock


\bibitem[Jones and Karsten(2008)]%
        {jones2008giddens}
\bibfield{author}{\bibinfo{person}{Matthew~R Jones} {and} \bibinfo{person}{Helena Karsten}.} \bibinfo{year}{2008}\natexlab{}.
\newblock \showarticletitle{Giddens's structuration theory and information systems research}.
\newblock \bibinfo{journal}{\emph{MIS quarterly}} (\bibinfo{year}{2008}), \bibinfo{pages}{127--157}.
\newblock


\bibitem[Jovanovic and Campbell(2022)]%
        {jovanovic22}
\bibfield{author}{\bibinfo{person}{Mladan Jovanovic} {and} \bibinfo{person}{Mark Campbell}.} \bibinfo{year}{2022}\natexlab{}.
\newblock \showarticletitle{Generative artificial intelligence: Trends and prospects}.
\newblock \bibinfo{journal}{\emph{Computer}} \bibinfo{volume}{55}, \bibinfo{number}{10} (\bibinfo{year}{2022}), \bibinfo{pages}{107--112}.
\newblock


\bibitem[Kalliamvakou(2024)]%
        {kalliamvakou24}
\bibfield{author}{\bibinfo{person}{Eirini Kalliamvakou}.} \bibinfo{year}{2024}\natexlab{}.
\newblock \bibinfo{title}{A developer’s second brain: Reducing complexity through partnership with AI}.
\newblock


\bibitem[Khojah et~al\mbox{.}(2024)]%
        {khojah24}
\bibfield{author}{\bibinfo{person}{Ranim Khojah}, \bibinfo{person}{Mazen Mohamad}, \bibinfo{person}{Philipp Leitner}, {and} \bibinfo{person}{Francisco~Gomes de Oliveira~Neto}.} \bibinfo{year}{2024}\natexlab{}.
\newblock \showarticletitle{Beyond code generation: An observational study of chatgpt usage in software engineering practice}.
\newblock \bibinfo{journal}{\emph{Proceedings of the ACM on Software Engineering}} \bibinfo{volume}{1}, \bibinfo{number}{FSE} (\bibinfo{year}{2024}), \bibinfo{pages}{1819--1840}.
\newblock


\bibitem[Klein and Kozlowski(2000)]%
        {klein2000multilevel}
\bibfield{author}{\bibinfo{person}{Katherine~J Klein} {and} \bibinfo{person}{Steve~WJ Kozlowski}.} \bibinfo{year}{2000}\natexlab{}.
\newblock \showarticletitle{A multilevel approach to theory and research in organizations: Contextual, temporal, and emergent processes}.
\newblock \bibinfo{journal}{\emph{Multilevel theory, research, and methods in organizations: Foundations, extensions, and new directions}} (\bibinfo{year}{2000}), \bibinfo{pages}{3--90}.
\newblock


\bibitem[Ko(2019)]%
        {ko19}
\bibfield{author}{\bibinfo{person}{Amy~J Ko}.} \bibinfo{year}{2019}\natexlab{}.
\newblock \showarticletitle{Why we should not measure productivity}.
\newblock In \bibinfo{booktitle}{\emph{Rethinking Productivity in Software Engineering}}. \bibinfo{publisher}{Springer}.
\newblock


\bibitem[Korzy{\'n}ski et~al\mbox{.}(2024)]%
        {korzynski2024trust}
\bibfield{author}{\bibinfo{person}{Pawe{\l} Korzy{\'n}ski}, \bibinfo{person}{Susana Costa~e Silva}, \bibinfo{person}{Anna~Maria G{\'o}rska}, {and} \bibinfo{person}{Grzegorz Mazurek}.} \bibinfo{year}{2024}\natexlab{}.
\newblock \showarticletitle{Trust in AI and top management support in generative-AI adoption}.
\newblock \bibinfo{journal}{\emph{Journal of Computer Information Systems}} (\bibinfo{year}{2024}), \bibinfo{pages}{1--15}.
\newblock


\bibitem[Kuhail et~al\mbox{.}(2024)]%
        {kuhail2024will}
\bibfield{author}{\bibinfo{person}{Mohammad~Amin Kuhail}, \bibinfo{person}{Sujith~Samuel Mathew}, \bibinfo{person}{Ashraf Khalil}, \bibinfo{person}{Jose Berengueres}, {and} \bibinfo{person}{Syed Jawad~Hussain Shah}.} \bibinfo{year}{2024}\natexlab{}.
\newblock \showarticletitle{“Will I be replaced?” Assessing ChatGPT's effect on software development and programmer perceptions of AI tools}.
\newblock \bibinfo{journal}{\emph{Science of Computer Programming}}  \bibinfo{volume}{235} (\bibinfo{year}{2024}), \bibinfo{pages}{103111}.
\newblock


\bibitem[Kurup and Gupta(2022)]%
        {kurup2022factors}
\bibfield{author}{\bibinfo{person}{Sreejith Kurup} {and} \bibinfo{person}{Vivek Gupta}.} \bibinfo{year}{2022}\natexlab{}.
\newblock \showarticletitle{Factors influencing the AI adoption in organizations}.
\newblock \bibinfo{journal}{\emph{Metamorphosis}} \bibinfo{volume}{21}, \bibinfo{number}{2} (\bibinfo{year}{2022}), \bibinfo{pages}{129--139}.
\newblock


\bibitem[Lambiase et~al\mbox{.}(2024)]%
        {lambiase2024investigating}
\bibfield{author}{\bibinfo{person}{Stefano Lambiase}, \bibinfo{person}{Gemma Catolino}, \bibinfo{person}{Fabio Palomba}, \bibinfo{person}{Filomena Ferrucci}, {and} \bibinfo{person}{Daniel Russo}.} \bibinfo{year}{2024}\natexlab{}.
\newblock \showarticletitle{Investigating the role of cultural values in adopting large language models for software engineering}.
\newblock \bibinfo{journal}{\emph{ACM Transactions on Software Engineering and Methodology}} (\bibinfo{year}{2024}).
\newblock


\bibitem[Lambiase et~al\mbox{.}(2025)]%
        {lambiase25}
\bibfield{author}{\bibinfo{person}{Stefano Lambiase}, \bibinfo{person}{Gemma Catolino}, \bibinfo{person}{Fabio Palomba}, \bibinfo{person}{Filomena Ferrucci}, {and} \bibinfo{person}{Daniel Russo}.} \bibinfo{year}{2025}\natexlab{}.
\newblock \showarticletitle{Exploring Individual Factors in the Adoption of LLMs for Specific Software Engineering Tasks}.
\newblock \bibinfo{journal}{\emph{arXiv preprint arXiv:2504.02553}} (\bibinfo{year}{2025}).
\newblock


\bibitem[Lee et~al\mbox{.}(2021)]%
        {lee21}
\bibfield{author}{\bibinfo{person}{Shoo~K Lee}, \bibinfo{person}{Sukhy~K Mahl}, {and} \bibinfo{person}{Brian~H Rowe}.} \bibinfo{year}{2021}\natexlab{}.
\newblock \showarticletitle{The Induced Productivity Decline hypothesis: more physicians, higher compensation and fewer services}.
\newblock \bibinfo{journal}{\emph{Healthcare Policy}} (\bibinfo{year}{2021}).
\newblock


\bibitem[Lewis(2015)]%
        {lewis15}
\bibfield{author}{\bibinfo{person}{Sarah Lewis}.} \bibinfo{year}{2015}\natexlab{}.
\newblock \showarticletitle{Qualitative inquiry and research design: Choosing among five approaches}.
\newblock \bibinfo{journal}{\emph{Health promotion practice}} \bibinfo{volume}{16}, \bibinfo{number}{4} (\bibinfo{year}{2015}), \bibinfo{pages}{473--475}.
\newblock


\bibitem[Li et~al\mbox{.}(2024)]%
        {li24b}
\bibfield{author}{\bibinfo{person}{Ze~Shi Li}, \bibinfo{person}{Nowshin~Nawar Arony}, \bibinfo{person}{Ahmed~Musa Awon}, \bibinfo{person}{Daniela Damian}, {and} \bibinfo{person}{Bowen Xu}.} \bibinfo{year}{2024}\natexlab{}.
\newblock \showarticletitle{AI tool use and adoption in software development by individuals and organizations: a grounded theory study}.
\newblock \bibinfo{journal}{\emph{arXiv preprint arXiv:2406.17325}} (\bibinfo{year}{2024}).
\newblock


\bibitem[Liang et~al\mbox{.}(2024)]%
        {liang24}
\bibfield{author}{\bibinfo{person}{Jenny~T Liang}, \bibinfo{person}{Chenyang Yang}, {and} \bibinfo{person}{Brad~A Myers}.} \bibinfo{year}{2024}\natexlab{}.
\newblock \showarticletitle{A large-scale survey on the usability of ai programming assistants: Successes and challenges}. In \bibinfo{booktitle}{\emph{Proceedings of the 46th IEEE/ACM international conference on software engineering}}. \bibinfo{pages}{1--13}.
\newblock


\bibitem[Marcus and Sch{\"u}tz(2005)]%
        {marcus05}
\bibfield{author}{\bibinfo{person}{Bernd Marcus} {and} \bibinfo{person}{Astrid Sch{\"u}tz}.} \bibinfo{year}{2005}\natexlab{}.
\newblock \showarticletitle{Who are the people reluctant to participate in research? Personality correlates of four different types of nonresponse as inferred from self-and observer ratings}.
\newblock \bibinfo{journal}{\emph{Journal of personality}} (\bibinfo{year}{2005}).
\newblock


\bibitem[Mayer et~al\mbox{.}(2025)]%
        {mayer25}
\bibfield{author}{\bibinfo{person}{Hannah Mayer}, \bibinfo{person}{Michael Chui}, {and} \bibinfo{person}{Roger Roberts}.} \bibinfo{year}{2025}\natexlab{}.
\newblock \bibinfo{booktitle}{\emph{Superagency in the workplace: Empowering people to unlock AI’s full potential}}.
\newblock \bibinfo{type}{{T}echnical {R}eport}. \bibinfo{institution}{McKinsey Digital}.
\newblock
\urldef\tempurl%
\url{https://www.mckinsey.com/capabilities/mckinsey-digital/our-insights/superagency-in-the-workplace-empowering-people-to-unlock-ais-full-potential-at-work}
\showURL{%
\tempurl}


\bibitem[Microsoft({[n.\,d.]})]%
        {microsoft24}
\bibfield{author}{\bibinfo{person}{Microsoft}.} \bibinfo{year}{[n.\,d.]}\natexlab{}.
\newblock \bibinfo{title}{AI at Work Is Here. Now Comes the Hard Part}.
\newblock \bibinfo{howpublished}{\url{https://www.microsoft.com/en-us/worklab/work-trend-index/ai-at-work-is-here-now-comes-the-hard-part}}.
\newblock
\newblock
\shownote{Accessed Mar.\ 2025}.


\bibitem[Miles et~al\mbox{.}(2014)]%
        {miles14}
\bibfield{author}{\bibinfo{person}{Matthew~B Miles}, \bibinfo{person}{A~Michael Huberman}, {and} \bibinfo{person}{Johnny Saldana}.} \bibinfo{year}{2014}\natexlab{}.
\newblock \bibinfo{booktitle}{\emph{Fundamentals of Qualitative Data Analysis}}.
\newblock \bibinfo{publisher}{Sage Los Angeles, CA}.
\newblock


\bibitem[Mills-Tettey et~al\mbox{.}(2007)]%
        {mills07}
\bibfield{author}{\bibinfo{person}{G~Ayorkor Mills-Tettey}, \bibinfo{person}{Anthony Stentz}, {and} \bibinfo{person}{M~Bernardine Dias}.} \bibinfo{year}{2007}\natexlab{}.
\newblock \showarticletitle{The dynamic hungarian algorithm for the assignment problem with changing costs}.
\newblock \bibinfo{journal}{\emph{Robotics Institute, Pittsburgh, PA, Tech. Rep. CMU-RI-TR-07-27}}  \bibinfo{volume}{7} (\bibinfo{year}{2007}).
\newblock


\bibitem[Newport(2021)]%
        {newport21}
\bibfield{author}{\bibinfo{person}{Cal Newport}.} \bibinfo{year}{2021}\natexlab{}.
\newblock \bibinfo{title}{The Frustration with Productivity Culture}.
\newblock \bibinfo{howpublished}{\url{https://www.newyorker.com/culture/office-space/the-frustration-with-productivity-culture}}.
\newblock
\newblock
\shownote{Accessed: 2025-06-04}.


\bibitem[Newquist(1994)]%
        {newquist1994brain}
\bibfield{author}{\bibinfo{person}{Harvey Newquist}.} \bibinfo{year}{1994}\natexlab{}.
\newblock \bibinfo{booktitle}{\emph{Brain Makers}}.
\newblock \bibinfo{publisher}{Editors \& Engineers, Limited}.
\newblock


\bibitem[Nowell et~al\mbox{.}(2017)]%
        {nowell17}
\bibfield{author}{\bibinfo{person}{Lorelli~S Nowell}, \bibinfo{person}{Jill~M Norris}, \bibinfo{person}{Deborah~E White}, {and} \bibinfo{person}{Nancy~J Moules}.} \bibinfo{year}{2017}\natexlab{}.
\newblock \showarticletitle{Thematic analysis: Striving to meet the trustworthiness criteria}.
\newblock \bibinfo{journal}{\emph{International journal of qualitative methods}} \bibinfo{volume}{16}, \bibinfo{number}{1} (\bibinfo{year}{2017}), \bibinfo{pages}{1609406917733847}.
\newblock


\bibitem[Orlikowski(1999)]%
        {orlikowski1999technologies}
\bibfield{author}{\bibinfo{person}{Wanda~Janina Orlikowski}.} \bibinfo{year}{1999}\natexlab{}.
\newblock \showarticletitle{Technologies-in-practice: an enacted lens for studying technology in organizations}.
\newblock  (\bibinfo{year}{1999}).
\newblock


\bibitem[Orlikowski(2000)]%
        {orlikowski2000using}
\bibfield{author}{\bibinfo{person}{Wanda~J Orlikowski}.} \bibinfo{year}{2000}\natexlab{}.
\newblock \showarticletitle{Using technology and constituting structures: A practice lens for studying technology in organizations}.
\newblock \bibinfo{journal}{\emph{Organization science}} \bibinfo{volume}{11}, \bibinfo{number}{4} (\bibinfo{year}{2000}), \bibinfo{pages}{404--428}.
\newblock


\bibitem[Orlikowski and Gash(1994)]%
        {orlikowski1994technological}
\bibfield{author}{\bibinfo{person}{Wanda~J Orlikowski} {and} \bibinfo{person}{Debra~C Gash}.} \bibinfo{year}{1994}\natexlab{}.
\newblock \showarticletitle{Technological frames: making sense of information technology in organizations}.
\newblock \bibinfo{journal}{\emph{ACM Transactions on Information Systems (TOIS)}} \bibinfo{volume}{12}, \bibinfo{number}{2} (\bibinfo{year}{1994}), \bibinfo{pages}{174--207}.
\newblock


\bibitem[Peng et~al\mbox{.}(2023)]%
        {peng23}
\bibfield{author}{\bibinfo{person}{Sida Peng}, \bibinfo{person}{Eirini Kalliamvakou}, \bibinfo{person}{Peter Cihon}, {and} \bibinfo{person}{Mert Demirer}.} \bibinfo{year}{2023}\natexlab{}.
\newblock \showarticletitle{The impact of ai on developer productivity: Evidence from github copilot}.
\newblock \bibinfo{journal}{\emph{arXiv preprint arXiv:2302.06590}} (\bibinfo{year}{2023}).
\newblock


\bibitem[Perman({[n.\,d.]})]%
        {perman10}
\bibfield{author}{\bibinfo{person}{Matt Perman}.} \bibinfo{year}{[n.\,d.]}\natexlab{}.
\newblock \bibinfo{title}{The Six Major Factors that Determine Knowledge Worker Productivity}.
\newblock \bibinfo{howpublished}{\url{https://www.whatsbestnext.com/2010/02/the-six-major-factors-that-determine-knowledge-worker-productivity/}}.
\newblock
\newblock
\shownote{Accessed Mar.\ 2025}.


\bibitem[Prasad~Agrawal(2024)]%
        {prasad2024towards}
\bibfield{author}{\bibinfo{person}{Kalyan Prasad~Agrawal}.} \bibinfo{year}{2024}\natexlab{}.
\newblock \showarticletitle{Towards adoption of generative AI in organizational settings}.
\newblock \bibinfo{journal}{\emph{Journal of Computer Information Systems}} \bibinfo{volume}{64}, \bibinfo{number}{5} (\bibinfo{year}{2024}), \bibinfo{pages}{636--651}.
\newblock


\bibitem[Rogelberg et~al\mbox{.}(2003)]%
        {rogelberg03}
\bibfield{author}{\bibinfo{person}{Steven~G Rogelberg} {et~al\mbox{.}}} \bibinfo{year}{2003}\natexlab{}.
\newblock \showarticletitle{Profiling active and passive nonrespondents to an organizational survey}.
\newblock \bibinfo{journal}{\emph{Jrnl. of Applied Psych.}} (\bibinfo{year}{2003}).
\newblock


\bibitem[Rogers et~al\mbox{.}(2014)]%
        {rogers2014diffusion}
\bibfield{author}{\bibinfo{person}{Everett~M Rogers}, \bibinfo{person}{Arvind Singhal}, {and} \bibinfo{person}{Margaret~M Quinlan}.} \bibinfo{year}{2014}\natexlab{}.
\newblock \showarticletitle{Diffusion of innovations}.
\newblock In \bibinfo{booktitle}{\emph{An integrated approach to communication theory and research}}. \bibinfo{publisher}{Routledge}, \bibinfo{pages}{432--448}.
\newblock


\bibitem[Roland and Shiman(2002)]%
        {roland2002strategic}
\bibfield{author}{\bibinfo{person}{Alex Roland} {and} \bibinfo{person}{Philip Shiman}.} \bibinfo{year}{2002}\natexlab{}.
\newblock \bibinfo{booktitle}{\emph{Strategic computing: DARPA and the quest for machine intelligence, 1983-1993}}.
\newblock \bibinfo{publisher}{MIT Press}.
\newblock


\bibitem[Russo(2024)]%
        {russo24}
\bibfield{author}{\bibinfo{person}{Daniel Russo}.} \bibinfo{year}{2024}\natexlab{}.
\newblock \showarticletitle{Navigating the complexity of generative ai adoption in software engineering}.
\newblock \bibinfo{journal}{\emph{ACM Transactions on Software Engineering and Methodology}} \bibinfo{volume}{33}, \bibinfo{number}{5} (\bibinfo{year}{2024}), \bibinfo{pages}{1--50}.
\newblock


\bibitem[Schmitz and Fulk(1991)]%
        {schmitz1991organizational}
\bibfield{author}{\bibinfo{person}{Joseph Schmitz} {and} \bibinfo{person}{Janet Fulk}.} \bibinfo{year}{1991}\natexlab{}.
\newblock \showarticletitle{Organizational colleagues, media richness, and electronic mail: A test of the social influence model of technology use}.
\newblock \bibinfo{journal}{\emph{Communication research}} \bibinfo{volume}{18}, \bibinfo{number}{4} (\bibinfo{year}{1991}), \bibinfo{pages}{487--523}.
\newblock


\bibitem[SciPy({[n.\,d.]})]%
        {spicyOptimize}
\bibfield{author}{\bibinfo{person}{SciPy}.} \bibinfo{year}{[n.\,d.]}\natexlab{}.
\newblock \bibinfo{title}{Optimization (scipy.optimize)}.
\newblock \bibinfo{howpublished}{\url{https://docs.scipy.org/doc/scipy/tutorial/optimize.html}}.
\newblock
\newblock
\shownote{Accessed Apr.\ 2025}.


\bibitem[Sergeyuk et~al\mbox{.}(2025)]%
        {sergeyuk2025human}
\bibfield{author}{\bibinfo{person}{Agnia Sergeyuk}, \bibinfo{person}{Ilya Zakharov}, \bibinfo{person}{Ekaterina Koshchenko}, {and} \bibinfo{person}{Maliheh Izadi}.} \bibinfo{year}{2025}\natexlab{}.
\newblock \showarticletitle{Human-AI Experience in Integrated Development Environments: A Systematic Literature Review}.
\newblock \bibinfo{journal}{\emph{arXiv preprint arXiv:2503.06195}} (\bibinfo{year}{2025}).
\newblock


\bibitem[Sichkarenko(2024)]%
        {sichkarenko24}
\bibfield{author}{\bibinfo{person}{Anastassiya Sichkarenko}.} \bibinfo{year}{2024}\natexlab{}.
\newblock \bibinfo{title}{The State of Developer Ecosystem 2024: The Unstoppable Rise of AI, Leading Languages, and Impact on Developer Experience}.
\newblock \bibinfo{howpublished}{\url{https://blog.jetbrains.com/team/2024/12/11/the-state-of-developer-ecosystem-2024-unveiling-current-developer-trends-the-unstoppable-rise-of-ai-adoption-leading-languages-and-impact-on-developer-experience/}}.
\newblock


\bibitem[Tang et~al\mbox{.}(2023)]%
        {tang2023empirical}
\bibfield{author}{\bibinfo{person}{Ningzhi Tang}, \bibinfo{person}{Meng Chen}, \bibinfo{person}{Zheng Ning}, \bibinfo{person}{Aakash Bansal}, \bibinfo{person}{Yu Huang}, \bibinfo{person}{Collin McMillan}, {and} \bibinfo{person}{T Li}.} \bibinfo{year}{2023}\natexlab{}.
\newblock \showarticletitle{An empirical study of developer behaviors for validating and repairing ai-generated code}. In \bibinfo{booktitle}{\emph{13th Workshop on the Intersection of HCI and PL}}.
\newblock


\bibitem[Tarafdar et~al\mbox{.}(2007)]%
        {tarafdar2007impact}
\bibfield{author}{\bibinfo{person}{Monideepa Tarafdar}, \bibinfo{person}{Qiang Tu}, \bibinfo{person}{Bhanu~S Ragu-Nathan}, {and} \bibinfo{person}{TS Ragu-Nathan}.} \bibinfo{year}{2007}\natexlab{}.
\newblock \showarticletitle{The impact of technostress on role stress and productivity}.
\newblock \bibinfo{journal}{\emph{Journal of management information systems}} \bibinfo{volume}{24}, \bibinfo{number}{1} (\bibinfo{year}{2007}), \bibinfo{pages}{301--328}.
\newblock


\bibitem[Taylor(2004)]%
        {taylor04}
\bibfield{author}{\bibinfo{person}{Frederick~Winslow Taylor}.} \bibinfo{year}{2004}\natexlab{}.
\newblock \bibinfo{booktitle}{\emph{Scientific management}}.
\newblock \bibinfo{publisher}{Routledge}.
\newblock


\bibitem[Toosi et~al\mbox{.}(2021)]%
        {toosi2021brief}
\bibfield{author}{\bibinfo{person}{Amirhosein Toosi}, \bibinfo{person}{Andrea~G Bottino}, \bibinfo{person}{Babak Saboury}, \bibinfo{person}{Eliot Siegel}, {and} \bibinfo{person}{Arman Rahmim}.} \bibinfo{year}{2021}\natexlab{}.
\newblock \showarticletitle{A brief history of AI: how to prevent another winter (a critical review)}.
\newblock \bibinfo{journal}{\emph{PET clinics}} \bibinfo{volume}{16}, \bibinfo{number}{4} (\bibinfo{year}{2021}), \bibinfo{pages}{449--469}.
\newblock


\bibitem[Uplevel(2024)]%
        {uplevel24}
\bibfield{author}{\bibinfo{person}{Uplevel}.} \bibinfo{year}{2024}\natexlab{}.
\newblock \bibinfo{booktitle}{\emph{AI Won't Solve Your Developer Productivity Problems for You}}.
\newblock \bibinfo{type}{{T}echnical {R}eport}. \bibinfo{institution}{Uplevel}.
\newblock
\urldef\tempurl%
\url{https://uplevelteam.com/blog/ai-for-developer-productivity}
\showURL{%
\tempurl}


\bibitem[Vendramin et~al\mbox{.}(2021)]%
        {vendramin21}
\bibfield{author}{\bibinfo{person}{Nelda Vendramin}, \bibinfo{person}{Giulia Nardelli}, {and} \bibinfo{person}{Christine Ipsen}.} \bibinfo{year}{2021}\natexlab{}.
\newblock \showarticletitle{Task-Technology Fit Theory: An approach for mitigating technostress}.
\newblock In \bibinfo{booktitle}{\emph{A handbook of theories on designing alignment between people and the office environment}}. \bibinfo{publisher}{Routledge}, \bibinfo{pages}{39--53}.
\newblock


\bibitem[Venkatesh et~al\mbox{.}(2003)]%
        {venkatesh2003user}
\bibfield{author}{\bibinfo{person}{Viswanath Venkatesh}, \bibinfo{person}{Michael~G Morris}, \bibinfo{person}{Gordon~B Davis}, {and} \bibinfo{person}{Fred~D Davis}.} \bibinfo{year}{2003}\natexlab{}.
\newblock \showarticletitle{User acceptance of information technology: Toward a unified view}.
\newblock \bibinfo{journal}{\emph{MIS quarterly}} (\bibinfo{year}{2003}), \bibinfo{pages}{425--478}.
\newblock


\bibitem[Venkatesh et~al\mbox{.}(2012)]%
        {venkatesh2012consumer}
\bibfield{author}{\bibinfo{person}{Viswanath Venkatesh}, \bibinfo{person}{James~YL Thong}, {and} \bibinfo{person}{Xin Xu}.} \bibinfo{year}{2012}\natexlab{}.
\newblock \showarticletitle{Consumer acceptance and use of information technology: extending the unified theory of acceptance and use of technology}.
\newblock \bibinfo{journal}{\emph{MIS quarterly}} (\bibinfo{year}{2012}), \bibinfo{pages}{157--178}.
\newblock


\bibitem[Wang et~al\mbox{.}(2024)]%
        {wang2024investigating}
\bibfield{author}{\bibinfo{person}{Ruotong Wang}, \bibinfo{person}{Ruijia Cheng}, \bibinfo{person}{Denae Ford}, {and} \bibinfo{person}{Thomas Zimmermann}.} \bibinfo{year}{2024}\natexlab{}.
\newblock \showarticletitle{Investigating and designing for trust in ai-powered code generation tools}. In \bibinfo{booktitle}{\emph{Proceedings of the 2024 ACM Conference on Fairness, Accountability, and Transparency}}. \bibinfo{pages}{1475--1493}.
\newblock


\bibitem[Weick(1990)]%
        {weick1990technology}
\bibfield{author}{\bibinfo{person}{Karl~E Weick}.} \bibinfo{year}{1990}\natexlab{}.
\newblock \showarticletitle{Technology as equivoque: Sensemaking in new technologies.}
\newblock  (\bibinfo{year}{1990}).
\newblock


\bibitem[Weisz et~al\mbox{.}(2023)]%
        {weisz2023toward}
\bibfield{author}{\bibinfo{person}{Justin~D Weisz}, \bibinfo{person}{Michael Muller}, \bibinfo{person}{Jessica He}, {and} \bibinfo{person}{Stephanie Houde}.} \bibinfo{year}{2023}\natexlab{}.
\newblock \showarticletitle{Toward general design principles for generative AI applications}.
\newblock \bibinfo{journal}{\emph{arXiv preprint arXiv:2301.05578}} (\bibinfo{year}{2023}).
\newblock


\bibitem[Weisz et~al\mbox{.}(2021)]%
        {weisz2021perfection}
\bibfield{author}{\bibinfo{person}{Justin~D Weisz}, \bibinfo{person}{Michael Muller}, \bibinfo{person}{Stephanie Houde}, \bibinfo{person}{John Richards}, \bibinfo{person}{Steven~I Ross}, \bibinfo{person}{Fernando Martinez}, \bibinfo{person}{Mayank Agarwal}, {and} \bibinfo{person}{Kartik Talamadupula}.} \bibinfo{year}{2021}\natexlab{}.
\newblock \showarticletitle{Perfection not required? Human-AI partnerships in code translation}. In \bibinfo{booktitle}{\emph{Proceedings of the 26th International Conference on Intelligent User Interfaces}}. \bibinfo{pages}{402--412}.
\newblock


\bibitem[Wu et~al\mbox{.}(2022)]%
        {wu22}
\bibfield{author}{\bibinfo{person}{Charley~M Wu}, \bibinfo{person}{Eric Schulz}, \bibinfo{person}{Timothy~J Pleskac}, {and} \bibinfo{person}{Maarten Speekenbrink}.} \bibinfo{year}{2022}\natexlab{}.
\newblock \showarticletitle{Time pressure changes how people explore and respond to uncertainty}.
\newblock \bibinfo{journal}{\emph{Scientific reports}} (\bibinfo{year}{2022}).
\newblock


\end{thebibliography}
